\documentclass[fleqn]{article}
\usepackage{longtable}
\usepackage{graphicx}

\begin {document}
%\pagenumbering{arabic}
%\frenchspacing
%parindent 0.0 cm
%\parskip 0.5cm
%\begin{center}
\begin{flushleft}
{\LARGE
{\bf Energy levels, radiative rates and electron impact excitation rates for transitions in He-like Ga XXX, \linebreak Ge XXXI, As XXXII, Se XXXIII and Br XXXIV}
}\\

\vspace{1.5 cm}

{\bf {Kanti  M  ~Aggarwal and Francis  P  ~Keenan}}\\ 

\vspace*{1.0cm}

Astrophysics Research Centre, School of Mathematics and Physics, Queen's University Belfast, Belfast BT7 1NN, Northern Ireland, UK\\ 

\vspace*{0.5 cm} 

e-mail: K.Aggarwal@qub.ac.uk \\

\vspace*{1.50cm}

Received  10 Sptember 2012\\
Accepted for publication 06 March 2013 \\
Published xx  Month 2013 \\
Online at stacks.iop.org/PhysScr/vol/number \\

\vspace*{1.5cm}

PACS Ref: 32.70 Cs, 34.80 Dp, 95.30 Ky

\vspace*{1.0 cm}

\hrule

\vspace{0.5 cm}
{\Large {\bf S}} This article has associated online supplementary data files \\
Tables 2 and 4 are available only in the electronic version at stacks.iop.org/PhysScr/vol/number

\end{flushleft}

\clearpage

%{\bf Abstract} 

\begin{abstract}

We report calculations of energy levels, radiative rates and electron impact excitation cross sections and rates for transitions in He-like Ga XXX,  Ge XXXI, As XXXII, Se XXXIII and Br XXXIV.  The {\sc grasp} (general-purpose relativistic atomic structure package) is adopted for calculating energy levels and radiative rates. For determining the collision strengths, and subsequently the excitation rates, the Dirac Atomic R-matrix Code ({\sc darc}) is used.  Oscillator strengths, radiative rates and line strengths are reported for all E1, E2, M1 and M2 transitions among the lowest 49 levels of each ion. Additionally, theoretical lifetimes are provided for all 49 levels of the above five ions. Collision strengths are averaged over a Maxwellian velocity distribution and the effective collision strengths  obtained listed over a wide temperature range up to 10$^{8}$ K. Comparisons are made with similar data obtained using the Flexible Atomic Code ({\sc fac}) to  highlight the importance of resonances, included in calculations with {\sc darc}, in the determination of effective collision strengths. Discrepancies between the collision strengths from {\sc darc} and {\sc fac}, particularly for some forbidden transitions, are also discussed.  Finally, discrepancies between the present results for effective collision strengths with the {\sc darc} code and earlier semi-relativistic  $R$-matrix data are noted over a wide range of electron temperatures for many transitions in all ions.

\end{abstract}

\clearpage

\section{Introduction}

Emission lines of He-like ions have been widely observed from a variety of astrophysical and laboratory plasmas. For example, lines of many He-like ions detected in solar plasmas at  x-ray wavelengths  (1--50 ${\rm \AA}$)  have been listed by Dere {\em et al} \cite{kpd01}, and in the regions 170--211 ${\rm \AA}$ and 245--291 ${\rm \AA}$ by Feldman {\em et al} \cite{uf}. Similarly, transitions from these ions have been observed in laboratory  plasmas \cite{bitter1} --  \cite{bitter3}.  Of particular interest are the resonance ($w$: 1s$^2$ $^1$S$_0$ -- 1s2p $^1$P$^{\rm o}_1$), intercombination ($x$ and $y$: 1s$^2$ $^1$S$_0$ -- 1s2p $^3$P$^{\rm o}_{2,1}$), and forbidden ($z$: 1s$^2$ $^1$S$_0 $ -- 1s2s $^3$S$_1$) lines, which are highly useful for  plasma diagnostics -- see, for example,  \cite{bk} -- \cite{pdg} and references therein.  However, to analyse observations, atomic data are required for a variety of parameters, such as energy levels, radiative rates ($A$- values), and excitation rates or equivalently the effective collision strengths ($\Upsilon$), which are obtained from the electron impact collision strengths ($\Omega$).  These data are also required for the modelling of fusion plasmas \cite{hpw}. Therefore, in a series of papers we have already reported atomic data for He-like ions up to $Z$=30  \cite{arxvii} -- \cite{fexxv} and for $Z$=36, i.e. Kr XXXV  \cite{kr35}. In this work we report similar data for the remaining five ions with  31 $\le Z \le$ 35, i.e. Ga XXX,  Ge XXXI, As XXXII, Se XXXIII and Br XXXIV.

The National Institute of Standards and Technology (NIST) have compiled and critically evaluated  energy levels for many He-like ions, and  have posted data on their website {\tt http://www.nist.gov/pml/data/asd.cfm}. Additionally, the $A$- value  for the 1s$^2$ $^1$S$_0$ -- 1s2s $^3$S$_1$ (1--2) magnetic dipole  transition of Ga XXX is available on the NIST website.  However, no collisional data are available in the literature for the ions under consideration. Therefore, in this paper we report a complete set of results (namely energy levels, radiative rates, lifetimes, and effective collision strengths) for all transitions among the lowest 49 fine-structure levels of these He-like ions, which belong to the 1s$^2$, 1s2$\ell$, 1s3$\ell$, 1s4$\ell$ and 1s5$\ell$ configurations. Finally, we also provide the $A$- values for four types of transitions, namely electric dipole (E1), electric quadrupole (E2), magnetic dipole (M1) and  magnetic quadrupole (M2), as these are required in  a complete plasma model.

For our calculations we employ the fully relativistic {\sc grasp} (general-purpose relativistic atomic structure  package) code for the determination of wavefunctions,  originally developed by Grant {\em et al} \cite{grasp0}  and subsequently revised by several workers, under the names {\sc grasp} \cite{grasp1},  {\sc grasp92} \cite{grasp2}, and {\sc grasp2k}  \cite{grasp2k}. However, the version adopted here is {\sc grasp0}, which is based on \cite{grasp0} and is revised by Dr P H Norrington.  This version contains most of the modifications undertaken in the other revised codes and is available on the website  {\tt http://web.am.qub.ac.uk/DARC/}.  {\sc grasp} is a fully relativistic code, and is based on the $jj$ coupling scheme.   Further relativistic corrections arising from the Breit interaction and QED effects (vacuum polarization and Lamb shift) have also been included in the same way as in the original version described in \cite{grasp0} and \cite{graspt}. Additionally, we have used the option of {\em extended average level} (EAL),  in which a weighted (proportional to 2$j$+1) trace of the Hamiltonian matrix is minimized. This produces a compromise set of orbitals describing closely lying states with  moderate accuracy. For our calculations of $\Omega$, we have adopted the {\em Dirac Atomic $R$-matrix Code} ({\sc darc}) of P H Norrington and I P Grant ({\tt http://web.am.qub.ac.uk/DARC/}). However, the code does not include the Breit and QED corrections, and hence the target energies obtained are slightly different (and comparatively less accurate) than from {\sc grasp}. Finally, for comparison purposes we have performed parallel calculations with the {\em Flexible Atomic Code} ({\sc fac}) of Gu \cite{fac}, available from the website {\tt {\verb+http://sprg.ssl.berkeley.edu/~mfgu/fac/+}}. This is also a fully relativistic code which provides a variety of atomic parameters, and (generally) yields results for energy levels and radiative rates comparable to {\sc grasp}  -- see, for example, \cite{tixxi} and  \cite{fe15}, and references therein. However, the differences in collision strengths and subsequently in effective collision strengths between {\sc fac} and  {\sc darc} can be large, particularly for  forbidden transitions, as demonstrated in  our earlier papers \cite{arxvii}--\cite{kr35}, and also discussed below in sections 5 and 6. Hence results from {\sc fac} will be helpful in assessing the accuracy of our energy levels and radiative rates, and in estimating the contribution of resonances to effective collision strengths, included in calculations with {\sc darc} but not in {\sc fac}.

\section{Energy levels}

The 1s$^2$, 1s2$\ell$, 1s3$\ell$, 1s4$\ell$ and 1s5$\ell$ configurations of He-like ions give rise to the lowest 49 levels listed in Table 1 (a--e), in which we compare our level energies with {\sc grasp} (obtained {\em without} and {\em with} the inclusion of Breit and QED effects) with the critically evaluated data compiled by NIST. Also included in these tables are our results obtained from the {\sc fac} code (FAC1), including the same CI (configuration interaction) as in {\sc grasp}.  Our level energies obtained without the Breit and QED effects (GRASP1) are consistently higher than the NIST values by  up to $\sim$ 1.7 Ryd,  similar to the effect observed for other He-like ions \cite{arxvii}--\cite{kr35}. However, the orderings are  nearly the same as  those of NIST. The inclusion of Breit and QED effects (GRASP2) lowers the energies by a maximum of  $\sim$ 1.9 Ryd, depending on the ion. As expected, the contribution of Breit and QED effects increases with increasing $Z$, but our GRASP2 energies are lower than the NIST listings by up to $\sim$ 0.4 Ryd, depending on the ion. In addition, the orderings have slightly altered in a few instances, see for example the 4f and 5f levels. However, the energy differences between these swapped levels are very small. It may be noted that NIST compilations are mostly based on a variety of theoretical works resulting in small differences with our listed energies. Our FAC1 level energies  are consistently higher by up to 0.2 Ryd than the GRASP2 results, for all ions, and hence are comparatively in better agreement with the NIST listings. The level orderings from FAC1 are also in general agreement with the calculations from {\sc grasp}, but differ in some instances, particularly for the $n$ = 5 levels. This is mainly because the non-degeneracy among the levels of the $n$ = 5 configurations is very small. A further inclusion of the 1s6$\ell$ configurations, labelled FAC2 calculations in Table 1 (a--e), makes no appreciable difference either in the magnitude or orderings of the levels. 

Finally, in Table 1 (a--e) we also list the energies of Whiteford {\em et al} \cite{icft}, which are obtained with the {\em AutoStructure} ({\sc as}) code of Badnell \cite{as}, and are available at the {\sc apap} (Atomic Processes for Astrophysical Plasmas) website: {\tt {\verb+http://amdpp.phys.strath.ac.uk/UK_APAP/+}}. The {\sc as} energies are higher by up to $\sim$ 2 Ryd, depending on the ion, than the corresponding reference values from NIST and  our  results from the {\sc fac} and {\sc grasp} codes. Differences between the NIST and our theoretical energy levels with the latter calculations increase with increasing $Z$, and are due to  the higher-order relativistic effects being neglected in the {\sc as} calculations. More importantly, the level orderings are slightly different, particularly for levels 42 and higher, and the {\sc as}  energies for some of the levels  of the 5g configuration are non-degenerate. However,  the energy differences among the  levels of a configuration are very small, as noted above. To conclude, we may state that overall there is  general agreement between our calculations with the {\sc fac} and {\sc grasp} codes for the energy levels of the He-like ions considered here, and there is no major discrepancy with the NIST listings. However, the NIST energies are not available for some levels, particularly  of the $n$ = 5 configurations, for which our energy levels either from the GRASP2 or FAC1 calculations should be adopted in modelling applications. For the remaining levels, the critically compiled listings of  NIST may be preferred.

\section{Radiative rates}

The absorption oscillator strength ($f_{ij}$) and radiative rate $A_{ji}$ (in s$^{-1}$) for a transition $i \to j$ are related by the following expression:

\begin{equation}
f_{ij} = \frac{mc}{8{\pi}^2{e^2}}{\lambda^2_{ji}} \frac{{\omega}_j}{{\omega}_i} A_{ji}
 = 1.49 \times 10^{-16} \lambda^2_{ji} (\omega_j/\omega_i) A_{ji}
\end{equation}
where $m$ and $e$ are the electron mass and charge, respectively, $c$ is the velocity of light,  $\lambda_{ji}$ is the transition energy/wavelength in $\rm \AA$, and $\omega_i$
and $\omega_j$ are the statistical weights of the lower ($i$) and upper ($j$) levels, respectively. Similarly, the oscillator strength $f_{ij}$ (dimensionless) and the line strength $S$ (in atomic unit, 1 a.u. = 6.460$\times$10$^{-36}$ cm$^2$ esu$^2$) are related by the  standard equations given below.

\begin{flushleft}
For the electric dipole (E1) transitions 
\end{flushleft} 
\begin{equation}
A_{ji} = \frac{2.0261\times{10^{18}}}{{{\omega}_j}\lambda^3_{ji}} S^{{\rm E1}} \hspace*{0.5 cm} {\rm and} \hspace*{0.5 cm} 
f_{ij} = \frac{303.75}{\lambda_{ji}\omega_i} S^{{\rm E1}}, \\
\end{equation}
\begin{flushleft}
for the magnetic dipole (M1) transitions  
\end{flushleft}
\begin{equation}
A_{ji} = \frac{2.6974\times{10^{13}}}{{{\omega}_j}\lambda^3_{ji}} S^{{\rm M1}} \hspace*{0.5 cm} {\rm and} \hspace*{0.5 cm}
f_{ij} = \frac{4.044\times{10^{-3}}}{\lambda_{ji}\omega_i} S^{{\rm M1}}, \\
\end{equation}
\begin{flushleft}
for the electric quadrupole (E2) transitions 
\end{flushleft}
\begin{equation}
A_{ji} = \frac{1.1199\times{10^{18}}}{{{\omega}_j}\lambda^5_{ji}} S^{{\rm E2}} \hspace*{0.5 cm} {\rm and} \hspace*{0.5 cm}
f_{ij} = \frac{167.89}{\lambda^3_{ji}\omega_i} S^{{\rm E2}}, 
\end{equation}

\begin{flushleft}
and for the magnetic quadrupole (M2) transitions 
\end{flushleft}
\begin{equation}
A_{ji} = \frac{1.4910\times{10^{13}}}{{{\omega}_j}\lambda^5_{ji}} S^{{\rm M2}} \hspace*{0.5 cm} {\rm and} \hspace*{0.5 cm}
f_{ij} = \frac{2.236\times{10^{-3}}}{\lambda^3_{ji}\omega_i} S^{{\rm M2}}. \\
\end{equation}

In Table 2 (a--e) we present transition energies/wavelengths ($\lambda$, in $\rm \AA$), radiative rates ($A_{ji}$, in s$^{-1}$), oscillator strengths ($f_{ij}$, dimensionless), and line strengths ($S$, in a.u.), in length  form (Babushkin gauge) only, for all 336 electric dipole (E1) transitions among the 49 levels of the He-like ions considered here. The {\em indices} used  to represent the lower and upper levels of a transition have already been defined in Table 1 (a--e). Similarly, there are 391 electric quadrupole (E2), 316  magnetic dipole (M1), and 410 magnetic quadrupole (M2) transitions among the 49 levels. However, for these transitions only the $A$-values are listed in Table 2, and the corresponding results for the $f$- or $S$- values can be easily obtained using Eqs. (1--5).

As stated earlier, no other $A$-values are available in the literature with which to compare our results.  Therefore, we have performed another calculation with the {\sc fac} code of Gu \cite{fac}.  For all  strong transitions (f $\ge$ 0.01), the $A$-values from {\sc grasp} and {\sc fac}, in the  Babushkin gauge, agree to better than 10\% for  the five ions.  Furthermore, for a majority of the strong E1 transitions (f $\ge$ 0.01) the length and velocity (Coulomb gauge) forms in our {\sc grasp} calculations agree within 10\%. However, the differences are larger for a few  transitions, which are among the degenerate levels of a configuration, such as  10--11 ($f  \sim$ 1.4$\times$10$^{-4}$), 24--25 ($f  \sim$ 8.5$\times$10$^{-6}$) and 27--29 ($f  \sim$ 1.9$\times$10$^{-7}$) in  Ga XXX. This is because their transition energy ($\Delta E$) is very small and hence a slight variation in $\Delta E$ has a considerable effect on the $A$-values. For a few such weaker transitions ($f  <$ 10$^{-3}$) the two forms of the $f$- value differ by  orders of magnitude, for all ions. Finally, as for the energy levels the effect of additional CI is negligible on the $A$- values, as results for all transitions agree within 10\% with those obtained with the  inclusion of the $n$ = 6 configurations. Therefore,  for almost all strong E1 transitions our radiative rates should be accurate to better than 10\%. However, for a few weaker transitions the accuracy is comparatively lower.

\section{Lifetimes}

The lifetime $\tau$ for a level $j$ is defined as follows:

\begin{equation}  {\tau}_j = \frac{1}{{\sum_{i}^{}} A_{ji}}.  
\end{equation} 
 Since this is a measurable parameter, it provides a check on the accuracy of the calculations. Therefore, in Table 1 (a--e) we have also listed our calculated lifetimes, which include the contributions from four types of transitions, i.e. E1, E2, M1 and M2.  Unfortunately to our knowledge no similar theoretical or experimental data are available with which  to compare our results. However, based on the accuracy assessment of our $A$- values, we expect our lifetimes to have the same level of accuracy.
 
\section{Collision strengths}

Collision strengths ($\Omega$) are related to the more commonly known  collision cross section ($\sigma_{ij}$, $\pi{a_0}^2$) by the following relationship:

\begin{equation}
\Omega_{ij}({\rm E}) = {k^2_i}\omega_i\sigma_{ij}({\rm E})
\end{equation}
where ${k^2_i}$ is the incident energy of the electron and $\omega_i$ is the statistical weight of the initial state. Results for collisional data are preferred 
in the form of $\Omega$ because it is a symmetric and dimensionless quantity.

For the computation of collision strengths $\Omega$, we have employed the {\em Dirac atomic $R$-matrix code} ({\sc darc}), which includes the relativistic effects in a
systematic way, in both the target description and the scattering model. It is based on the $jj$ coupling scheme, and uses the  Dirac-Coulomb Hamiltonian in the $R$-matrix
approach. The $R$-matrix radii adopted for Ga XXX,  Ge XXXI, As XXXII, Se XXXIII and Br XXXIV are 2.88, 2.72, 2.56, 2.40 and 2.24 au, respectively. For all five ions, 60  continuum orbitals have been included for each channel angular momentum in the expansion of the wavefunction, allowing us to compute $\Omega$ up to an energy of  2150, 2350, 2650, 3000 and 3400 Ryd for  Ga XXX,  Ge XXXI, As XXXII, Se XXXIII and Br XXXIV, respectively. These energy ranges are sufficient to calculate values of effective collision strengths $\Upsilon$ (see section 6)  up to T$_e$ = 10$^{8}$ K, appropriate for applications to the modelling of high temperature laboratory plasmas.  The maximum number of channels for a partial wave is 217, and the corresponding size of the Hamiltonian matrix is 13 076. To obtain convergence of  $\Omega$ for all transitions and at all energies, we have included all partial waves with angular momentum $J \le$ 40.5, although a larger number would have been  preferable for the convergence of some allowed transitions, especially at higher energies. However, to account for higher neglected partial waves, we have included a top-up, based on the Coulomb-Bethe approximation \cite{ab} for allowed transitions and geometric series for others.

For illustration, in Figs. 1--3 we show the variation of $\Omega$ with angular momentum $J$ for three transitions of Ga XXX, namely 2--5 (1s2s $^3$S$_1$ -- 1s2p $^3$P$^o_1$), 2--11 (1s2s $^3$S$_1$ -- 1s3p $^3$P$^o_1$), and 9--12 (1s3p $^3$P$^o_0$ -- 1s3p $^3$P$^o_2$),  and at three energies of 1000, 1400 and 1800 Ryd. Values of $\Omega$ have fully converged for all {\em resonance} transitions (including the allowed ones), plus a majority of the allowed transitions among the higher excited levels,  as shown in Fig. 2 for the 2--11 transition. It is also clear from Fig. 2 that the need to include a larger range of partial waves increases with increasing energy. However, values of $\Omega$ have not converged for those allowed transitions whose $\Delta E$ is very small (mainly within the same $n$ complex), as shown for the 2--5 transition in Fig. 1. Similarly, values of $\Omega$ have (almost) converged for all forbidden transitions, including those whose $\Delta E$ is very small, such as the 9--12 transition shown in Fig. 3. Therefore,  for the allowed transitions within the same $n$ complex, our wide range of partial waves is not sufficient for the convergence of $\Omega$, for which a top-up has been included as mentioned above, and has been found to be appreciable. 

In Table 3 (a--e) we list our values of $\Omega$ for resonance transitions of Ga XXX,  Ge XXXI, As XXXII, Se XXXIII and Br XXXIV at energies {\em above} thresholds. The  indices used  to represent the levels of a transition have already been defined in Table 1 (a--e). Unfortunately, no similar data are available for comparison purposes as already stated in section 1. Therefore, to assess the accuracy of  $\Omega$, we have performed another calculation using the {\sc fac} code of Gu \cite{fac}. This code is also fully relativistic, and is based on the well-known and widely-used {\em distorted-wave} (DW) method.  Furthermore, the same CI is included in {\sc fac} as in the calculations from {\sc darc}. Therefore, also included in Table 3 (a--e) for comparison purposes are the $\Omega$ values from {\sc fac} at a single {\em excited} energy E$_j$, which corresponds to an incident energy of $\sim$ 2000, 2100, 2300, 2400 and 2600 Ryd for Ga XXX,  Ge XXXI, As XXXII, Se XXXIII and Br XXXIV, respectively. For a majority of transitions the two sets of $\Omega$  generally agree well (within $\sim$ 20\%). However, the differences are larger for a few (particularly weaker) transitions. For example, for 70\% of the Ga XXX transitions, the values of $\Omega$ agree to within 20\% at an energy of 2000 Ryd, and the discrepancies for others are mostly within a factor of two, although for some transitions (such as: 19--49, 33--38/41/43/47/49),  the differences are up to an order of magnitude. However, most of these transitions are weak ($\Omega \sim$ 10$^{-6}$) and forbidden, i.e. the values of $\Omega$  have fully converged at {\em all} energies within our adopted range of partial waves in the calculations from the {\sc darc} code.  For such weak transitions, values of $\Omega$ from the {\sc fac} code are not assessed to be accurate.  The values of $\Omega$ are higher from {\sc darc} for some transitions, whereas for others are higher from the {\sc fac} code. Additionally, for a few transitions, such as 18--28/29/30/31,  values of $\Omega$ from the {\sc fac} code show an anomalous behaviour towards the higher end of the  energy range. This problem is common for all ions and  examples  can be seen in Fig. 6 of Aggarwal and Keenan \cite{mgxi} -- \cite{caxix}. The sudden anomalous behaviour in values of $\Omega$ from the {\sc fac} code is also responsible for the differences noted above for some of the transitions. Such anomalies for some transitions (both allowed and forbidden)  from the  {\sc fac} calculations  arise primarily because of the interpolation and extrapolation techniques employed in the  code, which is designed to generate a large amount of atomic data in a comparatively very short period of time, and without too large loss of accuracy. Similarly, some differences in  $\Omega$ are expected because the DW method generally overestimates the results due to the exclusion of channel coupling. Such discrepancies,  for a similar number of transitions, have also been noted for Ge XXXI, As XXXII, Se XXXIII and Br XXXIV. 

As a further comparison between the {\sc darc} and {\sc fac} values of $\Omega$, in Fig. 4 we show the variation of $\Omega$ with energy for three {\em allowed} transitions among the excited levels of Ga XXX, namely 5--14 (1s2p $^3$P$^o_1$ - 1s3d $^3$D$_2$), 12--26 (1s3p $^3$P$^o_2$ - 1s4d $^3$D$_3$), and 15--27 (1s3p $^1$P$^o_1$ -  1s4d $^1$D$_2$). Also included in this figure are the corresponding results obtained with the {\sc fac} code. For  many  transitions there are no discrepancies between the $f$- values obtained with the two different codes ({\sc grasp} and {\sc fac}), and therefore the values of $\Omega$ also agree to better than 20\%. However, the values of $\Omega$ obtained with {\sc fac}  are underestimated in comparison to  our calculations with {\sc darc},  particularly towards the lower end of the energy range. Similar comparisons between the two calculations with {\sc darc} and {\sc fac} are shown in Fig. 5 for three {\em forbidden} transitions of Ga XXX, namely 2--8 (1s2s $^3$S$_1$ - 1s3s $^3$S$_1$), 2--16 (1s2s $^3$S$_1$ - 1s3d $^3$D$_3$), and 6--12 (1s2p $^3$P$^o_2$ - 1s3p $^3$P$^o_2$). As in the case of  allowed transitions, for these forbidden ones  the agreement between the two calculations improves considerably with increasing energy, but the differences are significant towards the lower end of the energy range, as $\Omega$ from {\sc fac} are underestimated.   These anomalies are due to the interpolation and extrapolation techniques employed in the {\sc fac} code, as stated above. Therefore, on the basis of these and other comparisons discussed above, collision strengths from the {\sc fac} code are not assessed to be very accurate, over the entire energy range, for a majority of transitions for the above named five He-like ions. However, we do not see any apparent deficiency in our {\sc darc} calculations for $\Omega$, and estimate our results to be accurate to better than 20\% for a majority of the (strong) transitions. 

\section{Effective collision strengths}

Excitation rates, in addition to energy levels and radiative rates, are required for plasma modelling, and are determined from the collision strengths ($\Omega$). Since the
threshold energy region is dominated by numerous closed-channel (Feshbach) resonances, values of $\Omega$ need to be calculated in a fine energy mesh  to accurately
account for their contribution. Furthermore, in a plasma electrons have a wide distribution of velocities, and therefore values of $\Omega$ are generally averaged over a
{\em Maxwellian} distribution as follows:

\begin{equation}
\Upsilon(T_e) = \int_{0}^{\infty} {\Omega}(E) {\rm exp}(-E_j/kT_e) d(E_j/{kT_e}),
\end{equation}
where $k$ is Boltzmann constant, T$_e$ is electron temperature in K, and $E_j$ is the electron energy with respect to the final (excited) state. Once the value of $\Upsilon$ is
known the corresponding results for the excitation $q(i,j)$ and de-excitation $q(j,i)$ rates can be easily obtained from the following equations:

\begin{equation}
q(i,j) = \frac{8.63 \times 10^{-6}}{{\omega_i}{T_e^{1/2}}} \Upsilon {\rm exp}(-E_{ij}/{kT_e}) \hspace*{1.0 cm}{\rm cm^3s^{-1}}
\end{equation}
and
\begin{equation}
q(j,i) = \frac{8.63 \times 10^{-6}}{{\omega_j}{T_e^{1/2}}} \Upsilon \hspace*{1.0 cm}{\rm cm^3 s^{-1}},
\end{equation}
where $\omega_i$ and $\omega_j$ are the statistical weights of the initial ($i$) and final ($j$) states, respectively, and $E_{ij}$ is the transition energy. The contribution of resonances may enhance the values of $\Upsilon$ over those of the background  collision strengths ($\Omega_B$), especially for the forbidden transitions, by up to an order of magnitude (or even more) depending on the transition, and particularly at low temperatures.  Similarly, values of $\Omega$ need to be calculated over a wide energy range (above thresholds) in order to obtain convergence of the integral in Eq. (8), as demonstrated in Fig. 7 of Aggarwal and Keenan \cite{ni11a}. 

To delineate resonances, we have performed our calculations of $\Omega$ at up to  $\sim$ 127 500 energies in the thresholds region, depending on the ion. Close to thresholds ($\sim$0.1 Ryd above a threshold) the energy mesh is 0.001 Ryd, and away from thresholds is 0.002 Ryd. Thus care has been taken to include as many resonances as possible, and with as fine a resolution as is computationally feasible. The density and importance of resonances can be appreciated from Figs. 6--9, where we plot $\Omega$ as a function of energy in the thresholds region for the four most important transitions of He-like ions, namely 1--2 ($z$: 1s$^2$ $^1$S$_0$ -- 1s2s $^3$S$_1$), 1--5 ($y$: 1s$^2$ $^1$S$_0$ -- 1s2p $^3$P$^o_1$),  1--6 ($x$: 1s$^2$ $^1$S$_0$ -- 1s2p $^3$P$^o_2$), and 1--7 ($w$: 1s$^2$ $^1$S$_0$ -- 1s2p $^1$P$^o_1$). Resonances shown in these figures are for transitions in Ga XXX, but similar dense  resonances have been noted for  all He-like ions. For some transitions, such as 1--2, 1--5 and 1--6, the resonances are dense, particularly at energies just above the thresholds, and are spread over a wide  energy range of $\sim$ 150 Ryd. These near-threshold resonances affect the values of $\Upsilon$ particularly towards the lower end of the temperature range. 

Our calculated values of $\Upsilon$ are listed in Table 4 (a--e) over a wide temperature range up to 10$^{8}$ K, suitable for applications in a variety of plasmas. As stated in section 1, there are no other results available with which to compare. Therefore,  we have also calculated values of $\Upsilon$ from our non-resonant $\Omega$ data obtained with the {\sc fac} code. These calculations are particularly helpful in  providing an estimate of the importance of resonances in the determination of excitation rates. In Table 4 (a--e) we have included these results from {\sc fac} at the lowest and the highest calculated temperatures for Ga XXX,  Ge XXXI, As XXXII, Se XXXIII and Br XXXIV.  However, for discussion we focus solely on transitions in Ga XXX. At T$_e$ = 10$^{6.4}$ K, our resonance-resolved values of $\Upsilon$ are higher by over 20\% for about 37\% of the transitions. Generally, the differences for a majority of the transitions are within a factor of two, but are up to a factor of five  for a few.   Furthermore, for a majority of transitions the $\Upsilon$ values from the {\sc darc} code are higher,  partly due to  the inclusion of resonances in the calculations with {\sc darc} but also because of the underestimation of $\Omega$ at lower energies in the {\sc fac} calculations, as demonstrated in Figs. 4 and 5 and discussed in section 5.  In the case of the most important $w$, $x$, $y$, and $z$ lines, resonances have significantly enhanced the values of $\Upsilon$, by about a  factor of 2.5, for the $z$ (1--2: 1s$^2$ $^1$S$_0$ -- 1s2s $^3$S$_1$) transition. A similar comparison at the highest temperature of our calculations, i.e. 10$^{7.8}$ K, indicates that about 39\% of the transitions in Ga XXX show differences of over 20\% between the {\sc darc} and {\sc fac} values for $\Upsilon$. These differences are generally within a factor of two, but are higher (up to an order of magnitude) for a few, such as:  18--28 (allowed transition) and 18--29/30/31 (forbidden).  For all such transitions  the {\sc fac} results are {\em higher}, mainly because of the sudden anomalies in the values of $\Omega$, as discussed in section 5. Similar differences in  $\Upsilon$ are noted for a comparable number of transitions in the other He-like ions, as shown in Table 4 (a--e).

 Whiteford {\em et al} \cite{icft} have undertaken  semi-relativistic calculations for Ar XVII and Fe XXV, using the standard $R$-matrix code of Berrington {\em et al}\, \cite{rm}. Recently they have placed their results for $\Upsilon$ for He-like ions up to $Z$ = 36 on the {\sc apap} website: {\tt {\verb+http://amdpp.phys.strath.ac.uk/UK_APAP/+}}. In their calculations electron exchange was included for partial waves up to $J$ = 10.5, but was neglected for higher $J$ values. Furthermore, the calculations were performed in the $LS$ coupling scheme and the corresponding results for $\Omega$ and subsequently $\Upsilon$ for fine-structure transitions obtained using their intermediate coupling frame transformation ({\sc icft}) method. However, the data obtained by such procedures are generally comparable with our fully relativistic results from the {\sc darc} code, as has already been demonstrated in several papers -- see for example,  Liang and Badnell \cite{lb}. Furthermore, Whiteford {\em et al} included the effect of radiation damping. This can reduce the contribution of resonances to the determination of $\Upsilon$ for some transitions, such as 1s2p $^1$P$^o_1$ -- 1s3s $^1$S$_0$, particularly towards the lower end of the temperature range  (see their Fig. 4). However,  at temperatures relevant to plasma modelling the effect is negligible. Radiation damping  may be significant for highly charged ions, because radiative decay rates are large and compete with autoionisation rates  \cite{twg}--\cite{dpz}. However, similar $R$-matrix calculations performed by Delahaye {\em et al} \cite{dpz} for several He-like ions up to $Z$ = 20,  with the inclusion of radiation damping, show that their contribution (i.e. reduction in  values of $\Upsilon$) is appreciable only at lower temperatures (below 10$^6$ K), which should not be important in plasma modelling applications for such highly ionised species.  This has also been confirmed in calculations by  Whiteford {\em et al} \cite{icft}, who have shown that the resultant uncertainties in the excited level populations for the important $w$, $x$, $y$ and $z$  lines of Ar XVII and Fe XXV, at appropriate temperatures and densities, are under 10\%.  Finally, in a comparatively more recent calculations for Fe XXV and Kr XXXV using the the {\sc darc} code, Griffin and Ballance \cite{damp}  have concluded that the damping effect on excitation to the vast majority of levels for both ions is small. Specifically, the differences between the two sets of calculations (with and without damping) when averaged over six temperatures ranging from 1.25$\times$10$^6$ to 3.12$\times$10$^8$ K for the lowest 30 resonance transitions are only 4\% for Fe XXV and 4.4\% for Kr XXXV. Similarly, the corresponding differences for all 1176 transitions over a wider range of nine temperatures are only 1.1\% and 1.3\%, for the respective ions.  These differences are  much  smaller than the errors introduced by other parameters, such as inadequacies of partial waves, configuration interaction, energy mesh, and inclusion of a limited energy range. We can therefore confidently conclude that radiation damping is not important for the He-like  ions considered here.

We now compare our results of $\Upsilon$ with those of  Whiteford {\em et al} \cite{icft}, and for illustration  focus only on  transitions in Ga XXX.  Differences between the two sets of results are quite significant  (over 20\%) for many transitions, and throughout the temperature range of the calculations. To be specific, at the lowest common temperature of 1.8$\times$10$^5$ K, the two sets of $\Upsilon$ differ by over 20\% for $\sim$35\% of the transitions. For a majority of transitions, these differences are within a factor of two, and for some  our results are higher whereas for most the reverse is true. However, for a few transitions the differences are up to a factor of five, and for the  `elastic'   transitions, such as:  14--16/17, 16--17, 23--24/26/27, and 24--26/27, the discrepancies are up to two orders of magnitude, with the $\Upsilon$ values of Whiteford {\em et al} being invariably higher. Most of these  belong to the degenerate levels of a state/configuration, and hence have very small transition energies.   To demonstrate the differences between the two calculations,  in Fig. 10 we compare values of $\Upsilon$  for three transitions of Ga XXX, namely  14--16 (1s3d $^3$D$_{2}$ --  1s3d $^3$D$_{3}$), 16--17 (1s3d $^3$D$_{3}$ -- 1s3d $^1$D$_{2}$), and 24--26 (1s4d $^3$D$_{2}$ -- 1s4d $^3$D$_{3}$). The differences in the $\Upsilon$ values are {\em not} due to resonances, but arise from the limitation of the approach adopted by Whiteford {\em et al} \cite{icft}, as recently discussed and demonstrated by Bautista {\em et al}  \cite{feiii}.  Similar large discrepancies are also observed with the calculations of Whiteford {\em et al} \cite{ar16} for transitions in Li-like ions, as discussed and demonstrated by Aggarwal and Keenan \cite{lia},\cite{lib}. The problem in the $R$-matrix code adopted by Whiteford {\em et al} \cite{icft},\cite{ar16} has been identified and rectified by Liang and Badnell \cite{lb}. However, since the calculations of Whiteford {\em et al} \cite{icft},\cite{ar16} were  performed more than a decade ago, limitations in their data for He-like ions remain.

Differences between our data for $\Upsilon$ from {\sc darc} and those of Whiteford {\em et al} \cite{icft} are not confined to lower temperatures, but cover the entire range of temperatures. For example, at the highest common temperature of 4.5$\times$10$^7$ K, the two sets of $\Upsilon$ differ by over 20\% for $\sim$13\% of the transitions of Ga XXX. Hence, there is comparatively a better convergence of the results at higher temperatures, but the $\Upsilon$ values of Whiteford {\em et al} are invariably higher. The differences for most transitions are within a factor of two, but for some are up to an order of magnitude,  such as: 19--43/45, 29--31 and 33--43/45,  and all transitions with the lower levels I $\ge$ 46. Most of these transitions are {\em forbidden} and $\Omega$ for these have {\em converged} within our adopted partial waves range, as discussed in section 5. To demonstrate the differences, in Fig. 11 we compare our results of $\Upsilon$ from {\sc darc} with those of Whiteford {\em et al} \cite{icft} for three transitions of Ga XXX, namely  46--48 (1s5g $^3$G$_3$ -- 1s5g $^3$G$_5$), 47--48 (1s5g $^3$G$_{4}$ -- 1s5g $^3$G$_5$) and 48--49 (1s5g $^3$G$_{5}$ -- 1s5g $^1$G$_4$). Since the collision strengths of Whiteford {\em et al} are  overestimated for many transitions, particularly  at the lower energies,  the corresponding results for $\Upsilon$ are affected throughout the entire temperature range. To conclude, we may state that the $\Upsilon$ results of Whiteford {\em et al} \cite{icft} for transitions in Ga XXX and  other He-like ions should not be as accurate as those presented  here.

\section{Conclusions}

In this paper we have presented results for energy levels and  radiative rates for four types of transitions (E1, E2, M1 and M2) among the lowest 49 levels of Ga XXX,  Ge XXXI, As XXXII, Se XXXIII and Br XXXIV belonging to the $n \le$ 5 configurations. Additionally, lifetimes for all the calculated levels have been reported, although unfortunately no other theoretical or experimental results are available with which to compare.  Furthermore, based on a variety of comparisons among various calculations with the {\sc grasp} and {\sc fac} codes, our data for radiative rates, oscillator strengths, line strengths, and lifetimes are judged to be accurate to better than 10\% for a majority of the strong transitions (levels). Similarly, the accuracy of our results for collision strengths and effective collision strengths is estimated to be better than 20\% for most transitions.  We have considered a large range of partial waves to achieve convergence of $\Omega$ at all energies, included a wide energy range to accurately calculate  values of $\Upsilon$ up to T$_e$ = 10$^{8}$ K, and resolved resonances in a fine energy mesh to account for their contributions. Hence we see no apparent deficiency in our reported results. However, the present data for effective collision strengths for transitions involving the levels of the $n$ = 5 configurations may be improved slightly by the inclusion of the levels of the $n$ = 6 configurations. We believe the present set of complete results for radiative and excitation rates for transitions in five He-like ions should be useful for the modelling of a variety of plasmas. 

\section*{Acknowledgment}
KMA is grateful to  AWE Aldermaston for financial support.

%\newpage

%\end{document}

\clearpage

\begin{figure*}
%\resizebox{2.0\columnwidth}{!}{%
 \includegraphics[scale=0.70,angle=-90]{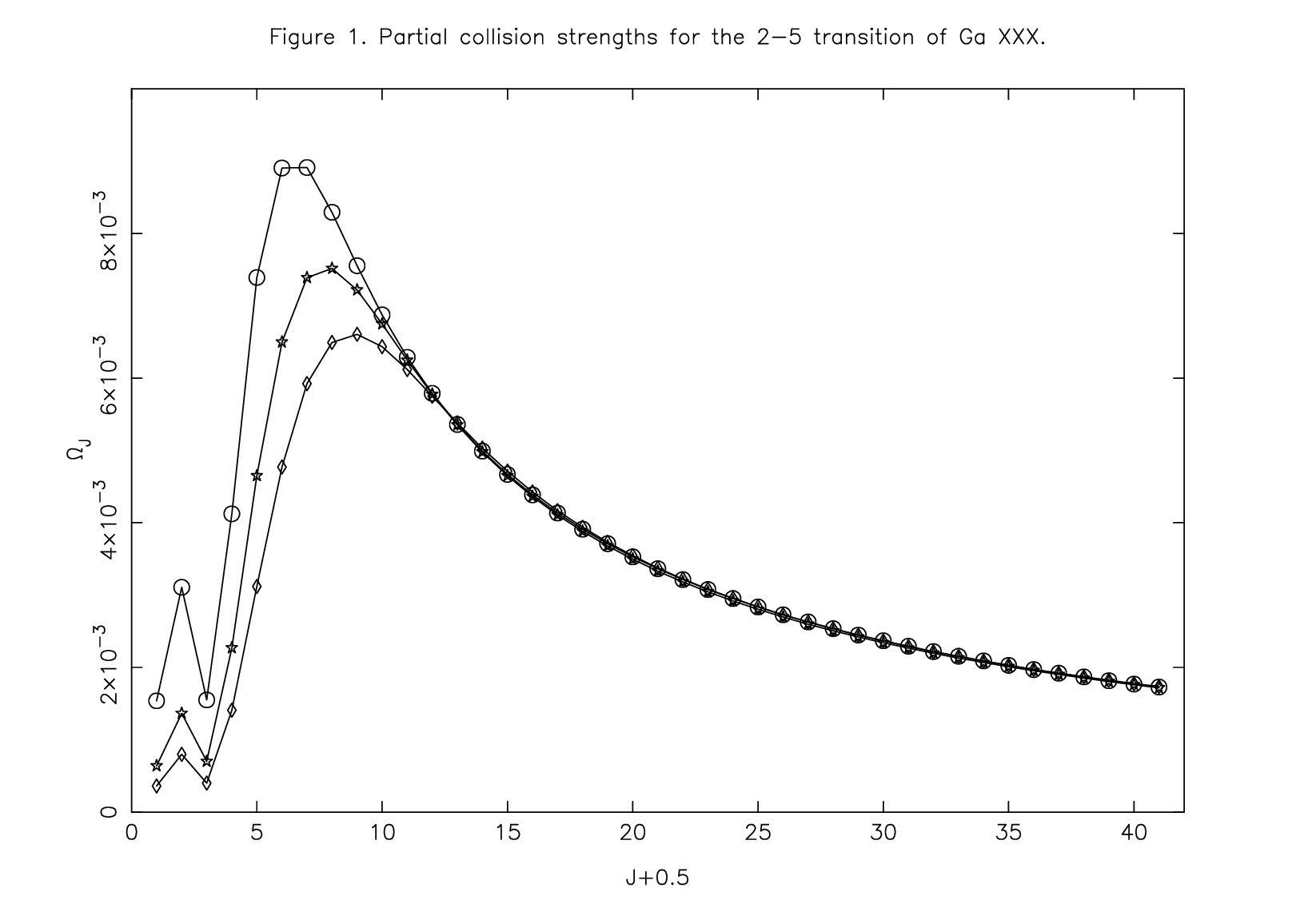}
%\vspace{5cm}       % Give the correct figure height in cm
\caption{Partial collision strengths for the 1s2s $^3$S$_1$ - 1s2p $^3$P$^o_1$ (2--5) transition of Ga XXX, 
at three energies of: 1000 Ryd (circles), 1400 Ryd (stars) and  1800 Ryd (diamonds).}
\label{fig:1}       % Give a unique label
\end{figure*}

\begin{figure*}
%\resizebox{2.0\columnwidth}{!}{%
\includegraphics[scale=0.70,angle=-90]{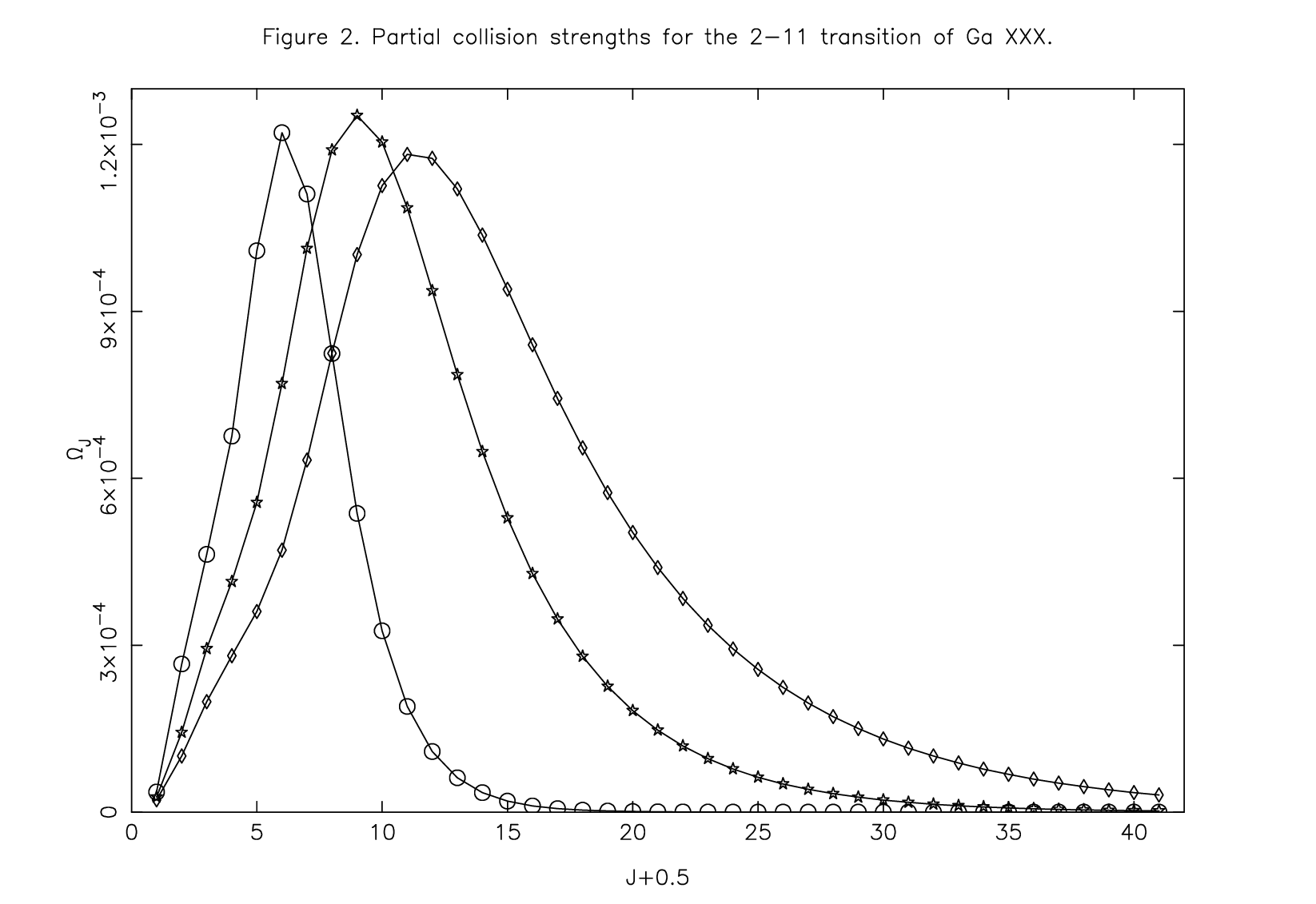}
%\vspace{5cm}       % Give the correct figure height in cm
\caption{Partial collision strengths for the 1s2s $^3$S$_1$ - 1s3p $^3$P$^o_1$ (2--11) transition of Ga XXX, 
at three energies of: 1000 Ryd (circles),  1400 Ryd (stars) and 1800 Ryd (diamonds).}
\label{fig:2}       % Give a unique label
\end{figure*}
\clearpage

\begin{figure*}
%\vspace*{5cm}       % Give the correct figure height in cm
%\resizebox{2.0\columnwidth}{!}{%
\includegraphics[scale=0.70,angle=-90]{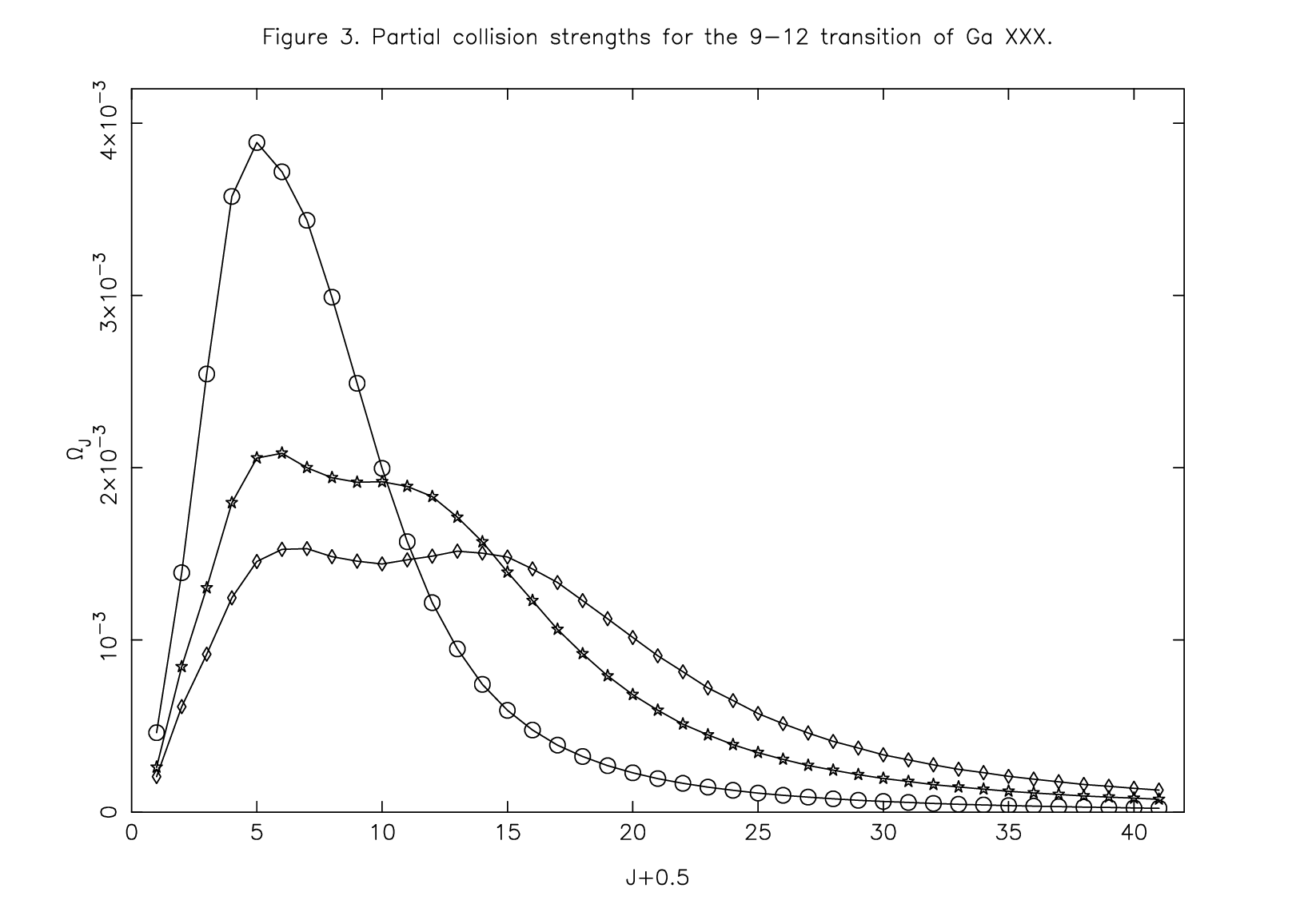}
 %}
\caption{Partial collision strengths for the 1s3p $^3$P$^o_0$ - 1s3p $^3$P$^o_2$ (9--12) transition of Ga XXX, 
at three energies of: 1000 Ryd (circles),  1400 Ryd (stars) and 1800 Ryd (diamonds).}
\label{fig:3}       % Give a unique label
\end{figure*}

\clearpage

\begin{figure*}
%\vspace*{5cm}       % Give the correct figure height in cm
%\resizebox{2.0\columnwidth}{!}{%
\includegraphics[scale=0.70,angle=-90]{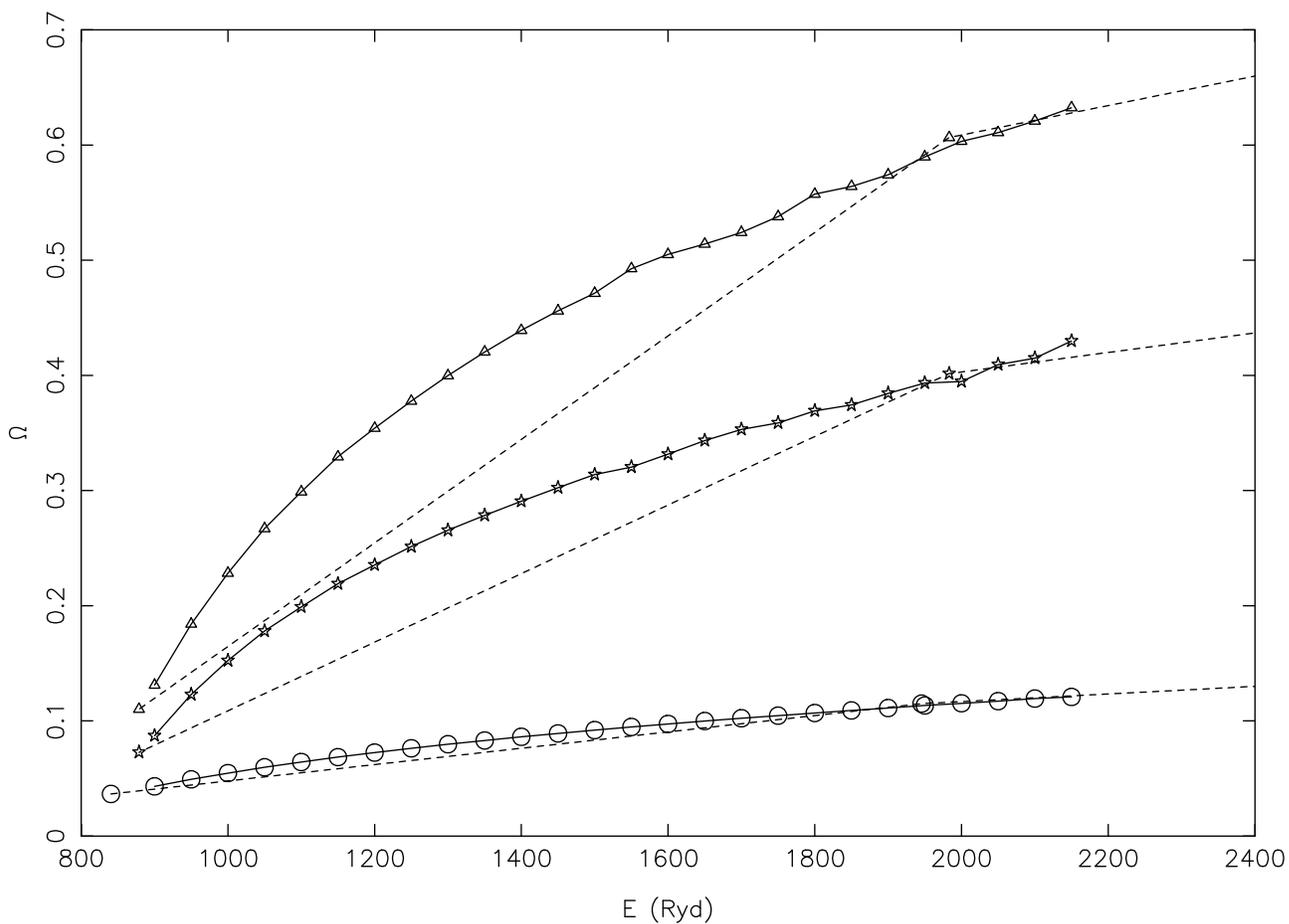}
%}
\caption{Comparison of collision strengths from our calculations from {\sc darc} (continuous curves) and {\sc fac} (broken curves) for the  5--14 (circles: 1s2p $^3$P$^o_1$ - 1s3d $^3$D$_2$), 12--26 (triangles : 1s3p $^3$P$^o_2$ - 1s4d $^3$D$_3$), and 15--27 (stars: 1s3p $^1$P$^o_1$ -  1s4d $^1$D$_2$) allowed transitions of Ga XXX.}
\label{fig:4}       % Give a unique label
\end{figure*}

\clearpage

\begin{figure*}
%\vspace*{5cm}       % Give the correct figure height in cm
%\resizebox{2.0\columnwidth}{!}{%
\includegraphics[scale=0.70,angle=-90]{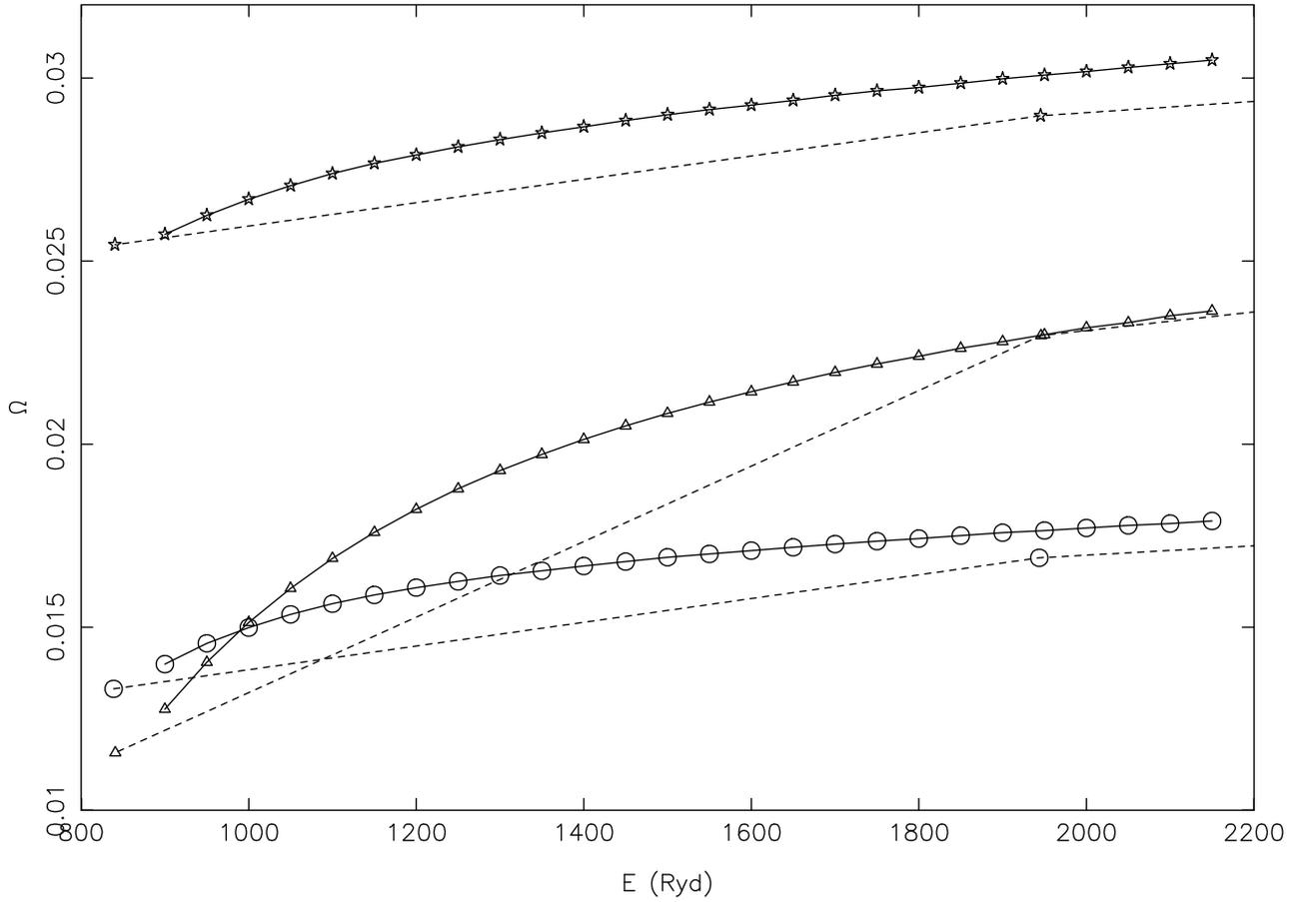}
%}
\caption{Comparison of collision strengths from our calculations from {\sc darc} (continuous curves) and {\sc fac} (broken curves) for the  2--8 (circles: 1s2s $^3$S$_1$ - 1s3s
$^3$S$_1$), 2--16 (triangles: 1s2s $^3$S$_1$ - 1s3d $^3$D$_3$), and 6--12 (stars: 1s2p $^3$P$^o_2$ - 1s3p $^3$P$^o_2$) forbidden transitions of Ga XXX.}
\label{fig:5}       % Give a unique label
\end{figure*}
\clearpage

\clearpage

\begin{figure*}
%\vspace*{5cm}       % Give the correct figure height in cm
%\resizebox{2.0\columnwidth}{!}{%
\includegraphics[scale=0.70,angle=90]{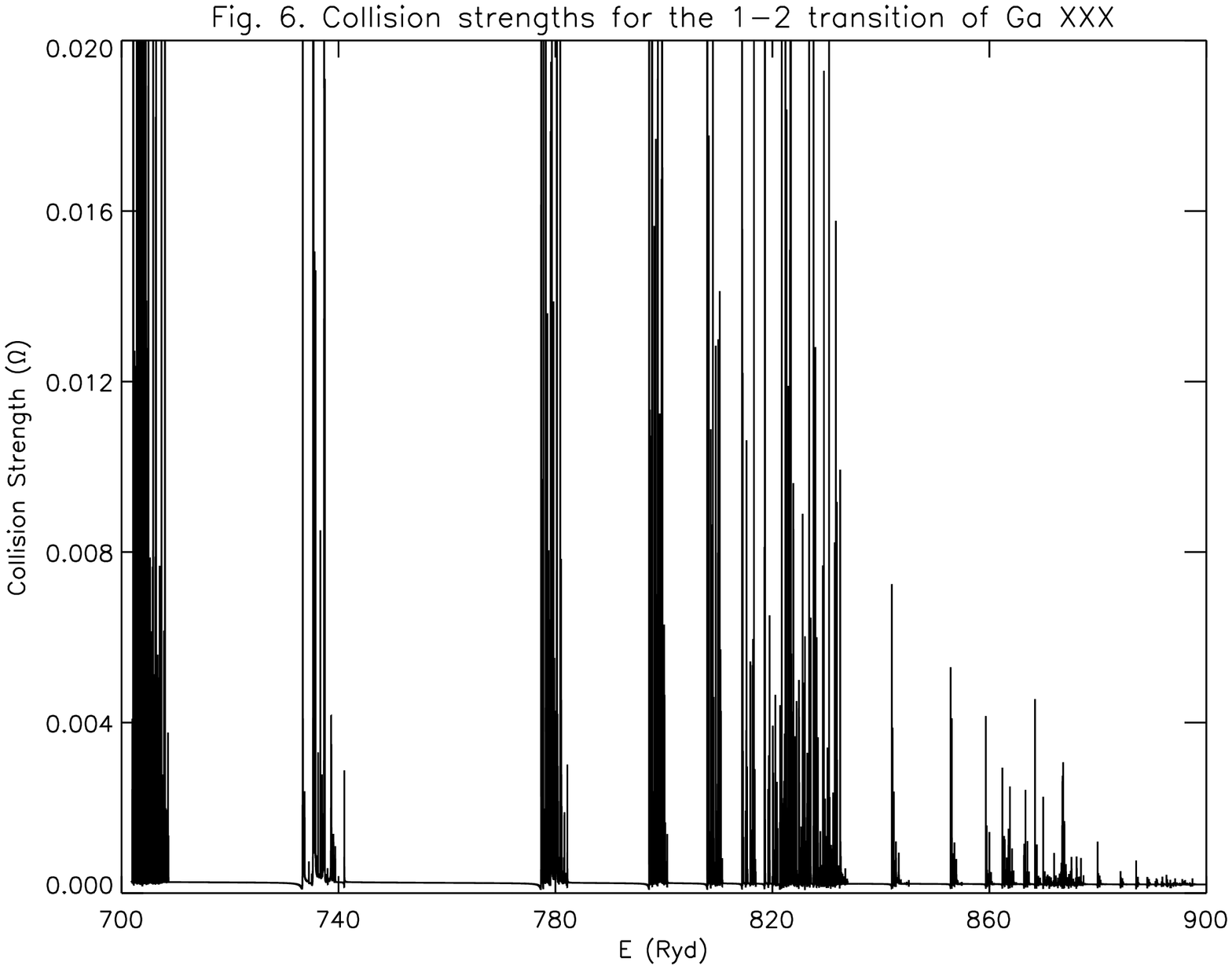}
%}
\caption{Collision strengths for the 1s$^2$ $^1$S$_0$ - 1s2s $^3$S$_1$   (1--2) transition of Ga XXX.}
\label{fig:7}       % Give a unique label
\end{figure*}

\begin{figure*}
%\vspace*{5cm}       % Give the correct figure height in cm
%\resizebox{2.0\columnwidth}{!}{%
\includegraphics[scale=0.70,angle=90]{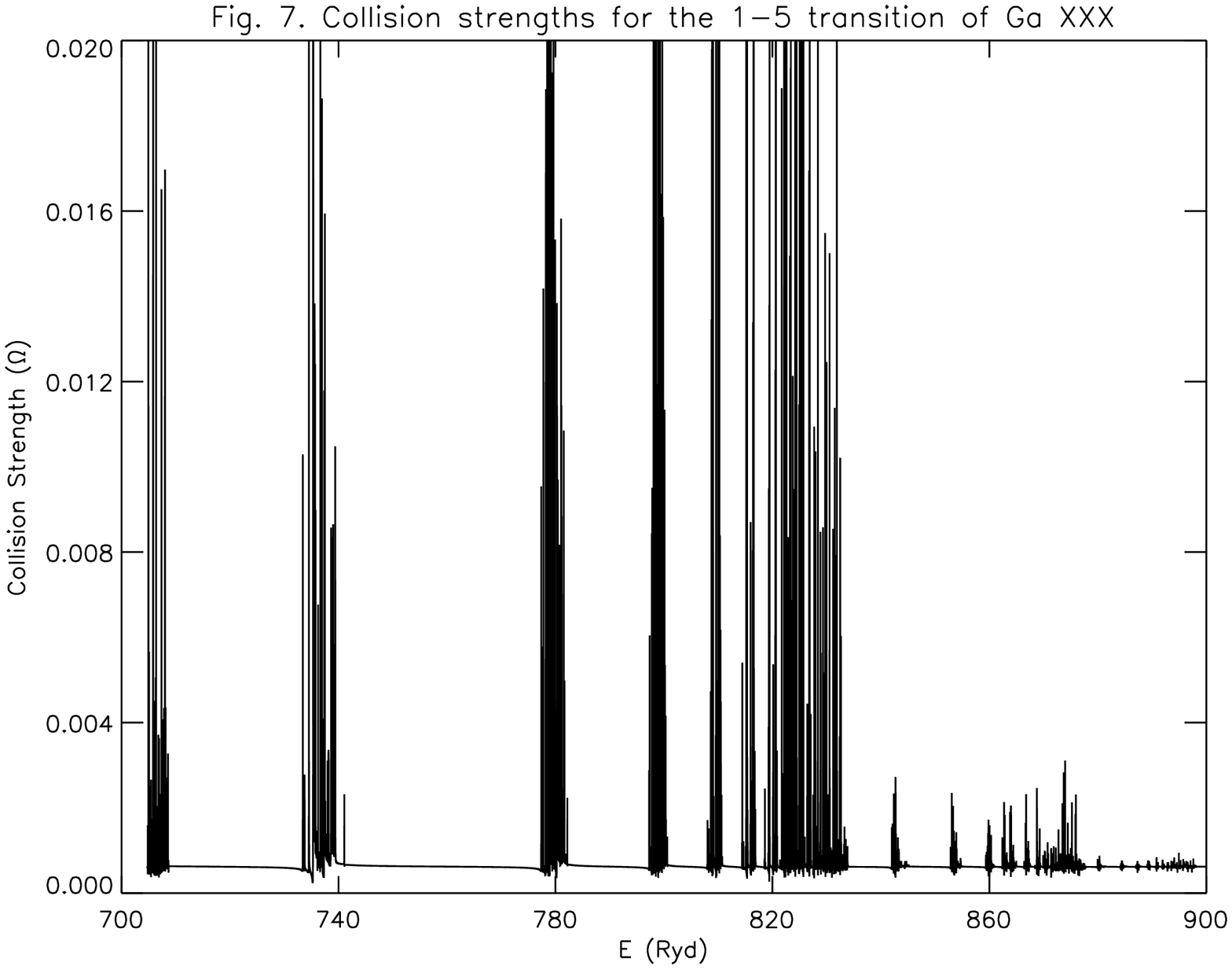}
%}
\caption{Collision strengths for the 1s$^2$ $^1$S$_0$ - 1s2p $^3$P$^o_1$ (1--5) transition of Ga XXX.}
\label{fig:8}       % Give a unique label
\end{figure*}

\begin{figure*}
%\vspace*{5cm}       % Give the correct figure height in cm
%\resizebox{2.0\columnwidth}{!}{%
\includegraphics[scale=0.70,angle=90]{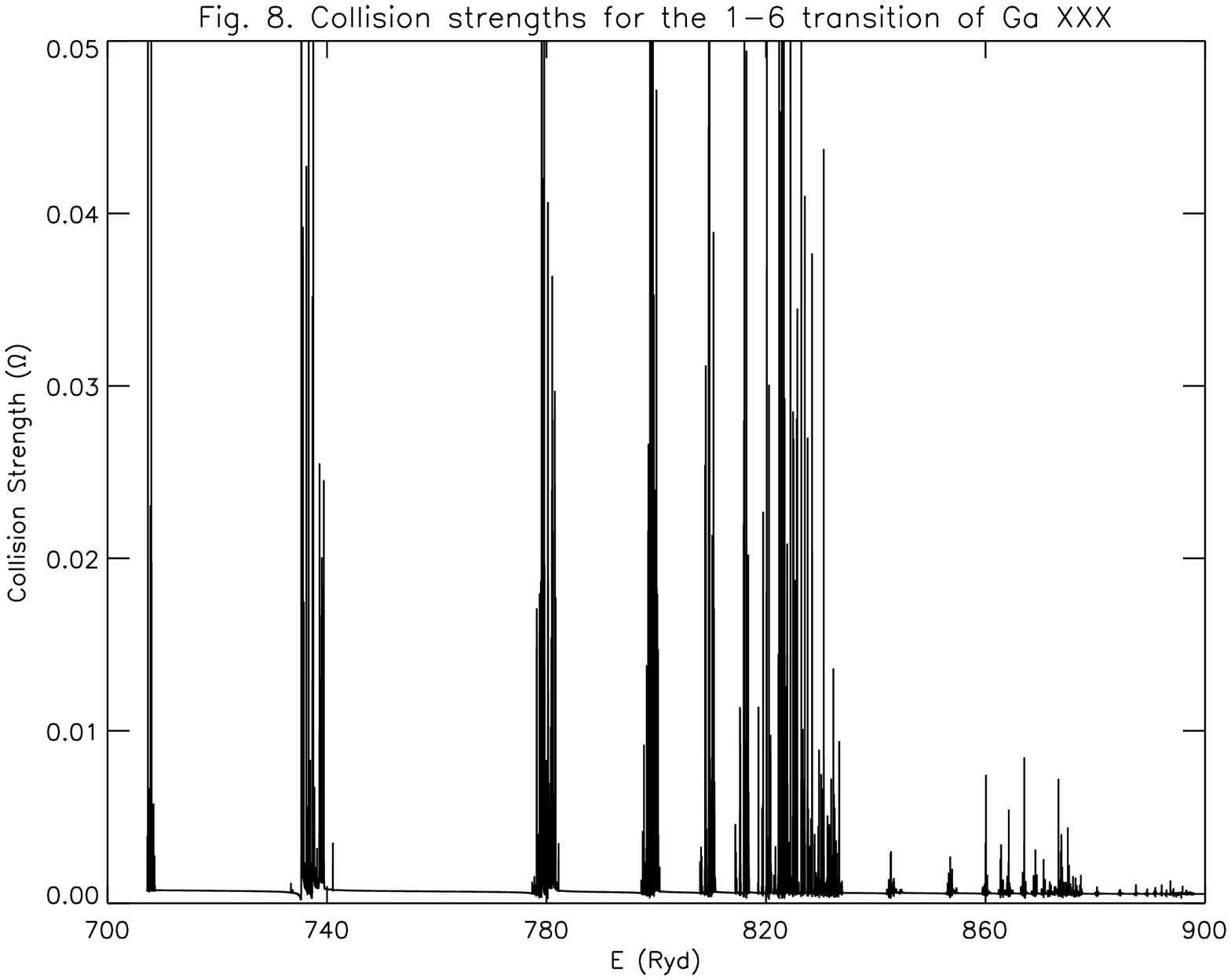}
%}
\caption{Collision strengths for the 1s$^2$ $^1$S$_0$ - 1s2p $^3$P$^o_2$ (1--6) transition of Ga XXX.}
\label{fig:9}       % Give a unique label
\end{figure*}

\begin{figure*}
%\vspace*{5cm}       % Give the correct figure height in cm
%\resizebox{2.0\columnwidth}{!}{%
\includegraphics[scale=0.70,angle=90]{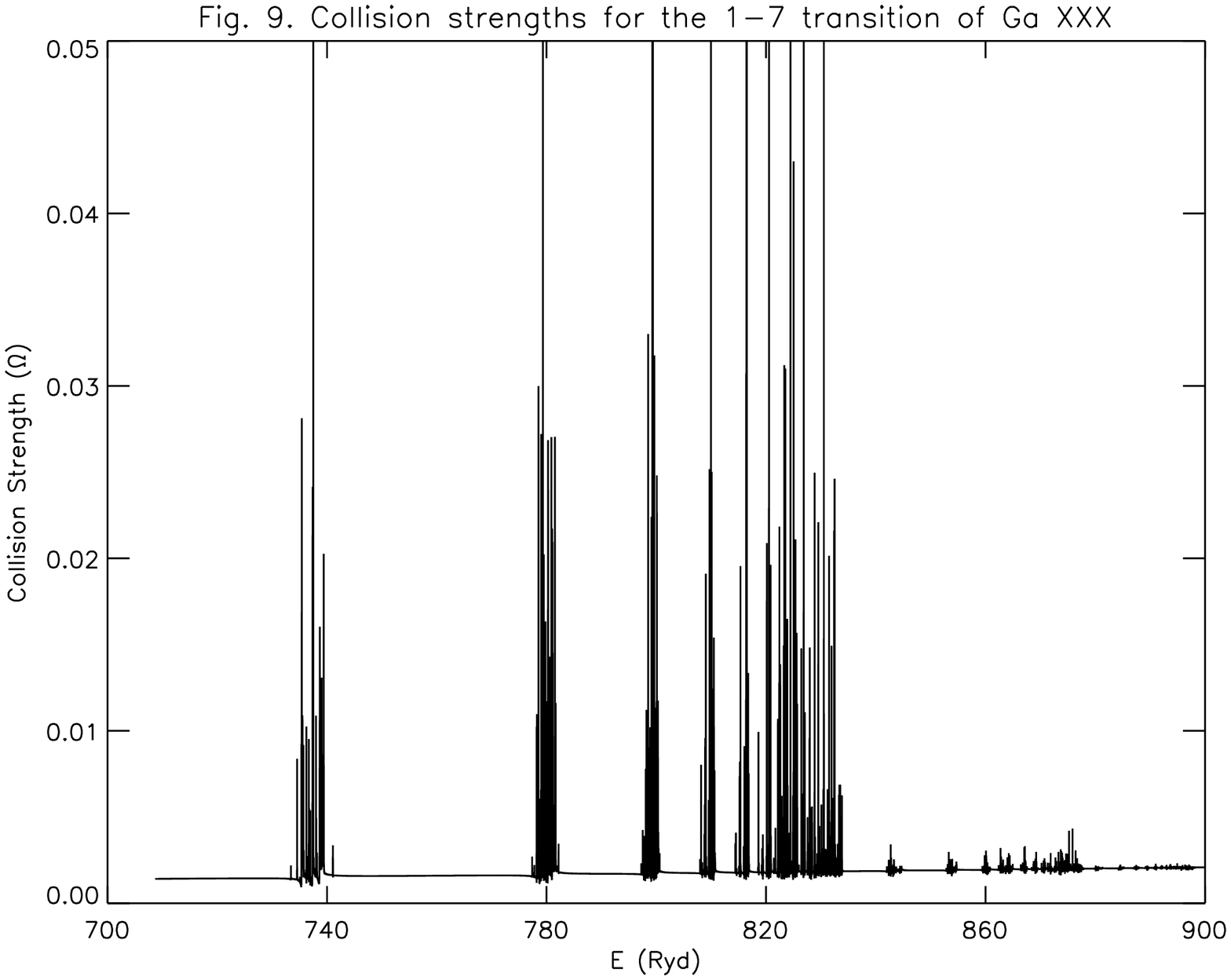}
%}
\caption{Collision strengths for the 1s$^2$ $^1$S$_0$ - 1s2p $^1$P$^o_1$ (1--7) transition of Ga XXX.}
\label{fig:10}       % Give a unique label
\end{figure*}

\setcounter{figure} {9}
\begin{figure*}
\includegraphics[scale=0.70,angle=-90]{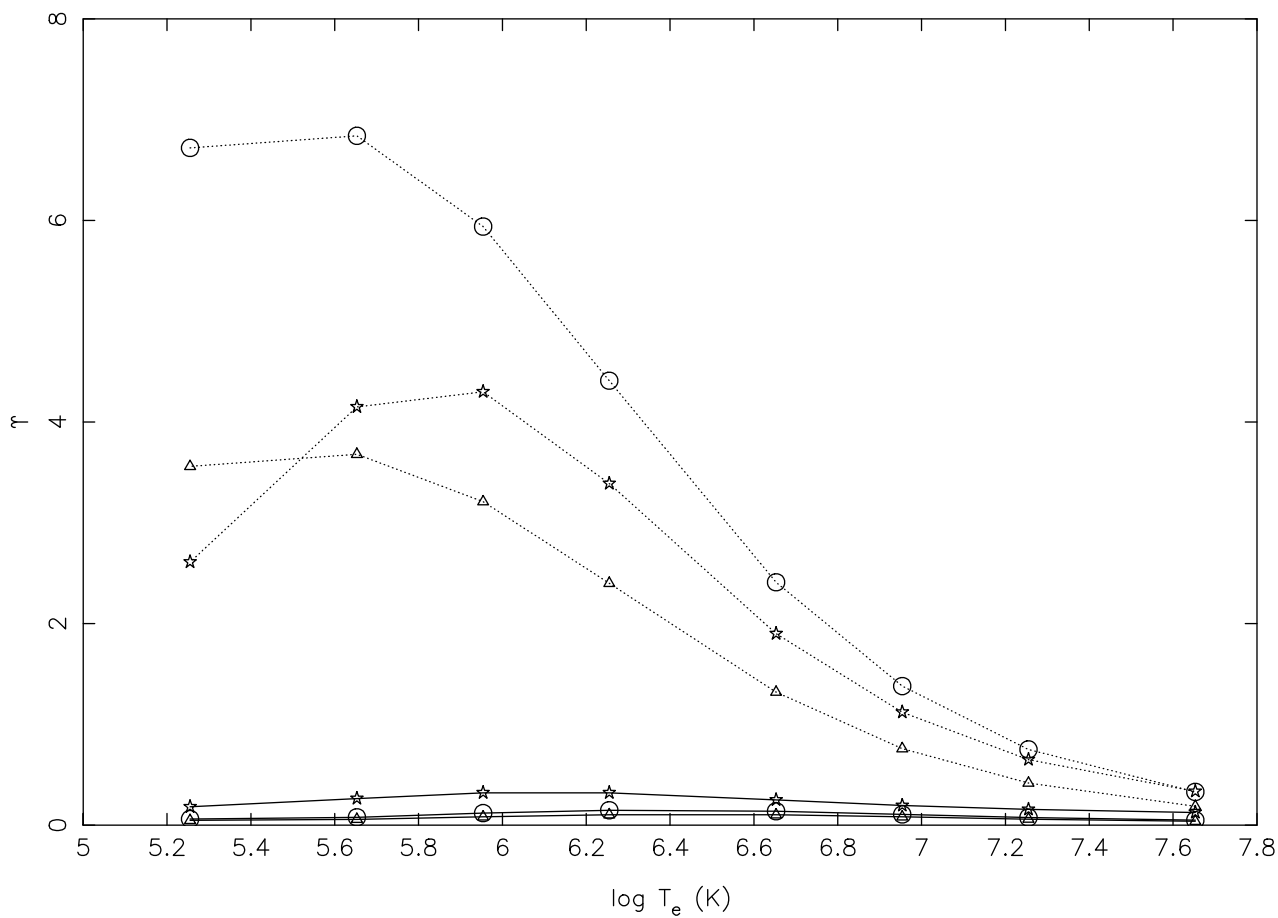}
\caption{Comparison of effective collision strengths for the 14--16 (circles: 1s3d $^3$D$_{2}$ --  1s3d $^3$D$_{3}$), 
16--17 (triangles: 1s3d $^3$D$_{3}$ -- 1s3d $^1$D$_{2}$), and 24--26 (stars: 1s4d $^3$D$_{2}$ -- 1s4d $^3$D$_{3}$)
 transitions of Ga XXX. Continuous and dotted curves are from the present {\sc darc} and earlier $R$- matrix codes \cite{icft}, respectively.}
\end{figure*}

\setcounter{figure} {10}
\begin{figure*}
\includegraphics[scale=0.70,angle=-90]{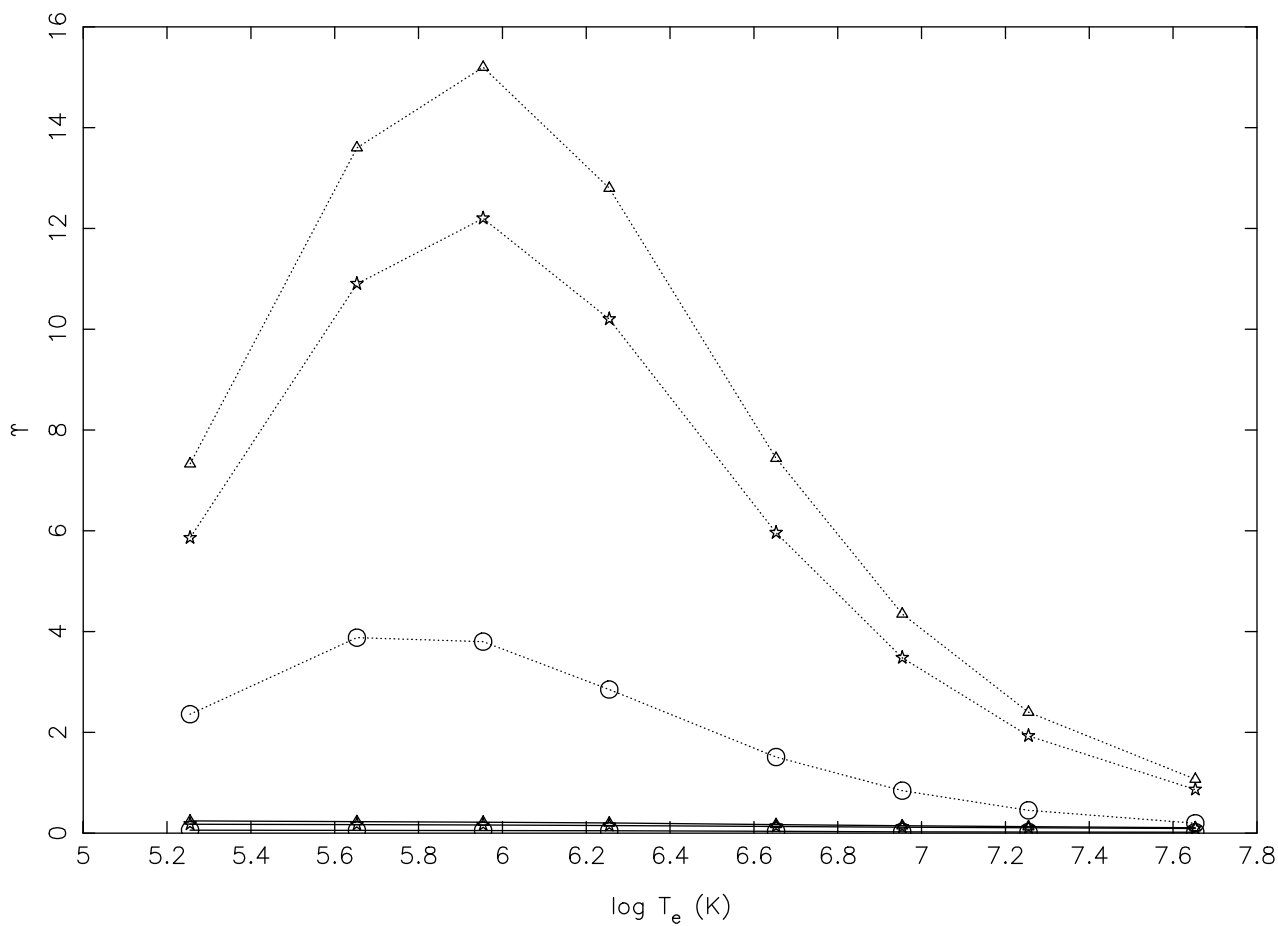}
\caption{Comparison of effective collision strengths for the 46--48 (circles: 1s5g $^3$G$_{3}$ --  1s5g $^3$G$_{5}$), 
47--48 (triangles: 1s5g $^3$G$_{4}$ -- 1s5g $^3$G$_{5}$), and 48--49 (stars: 1s5g $^3$G$_{5}$ -- 1s5g $^1$G$_{4}$)
 transitions of Ga XXX. Continuous and dotted curves are from the present {\sc darc} and earlier $R$- matrix codes \cite{icft}, respectively.}
\end{figure*}

%\end{document}

\clearpage

\begin{table*} 
\caption{a. Energy levels (in Ryd) of Ga XXX and their lifetimes ($\tau$, s). $a{\pm}b \equiv a{\times}$10$^{{\pm}b}$.} 
\begin{tabular}{rllrrrrrrr} \hline
Index  & \multicolumn{2}{c}{Configuration/Level} & NIST  & GRASP1 & GRASP2 &   FAC1  &    FAC2         &    AS        & $\tau$ (s)   \\    
\hline
   1  &   1s$^2$  &  $^1$S$_0  $ &    0.00000  &    0.00000  &     0.00000   &   0.00000  &   0.00000	&   0.00000   &  ........     \\
   2  &   1s2s    &  $^3$S$_1  $ &  700.86597  &  701.83875  &   700.59790   & 700.73132  & 700.73102	& 702.11621   &  8.137-10     \\
   3  &   1s2p    &  $^3$P$_0^o$ &  703.52441  &  704.29773  &   703.26678   & 703.43005  & 703.42999	& 704.47528   &  2.070-09     \\
   4  &   1s2s    &  $^1$S$_0  $ &  703.73382  &  704.60309  &   703.51764   & 703.66174  & 703.66138	& 704.94012   &  2.190-03     \\
   5  &   1s2p    &  $^3$P$_1^o$ &  703.71860  &  704.72906  &   703.46155   & 703.63263  & 703.63251	& 704.91608   &  6.376-15     \\
   6  &   1s2p    &  $^3$P$_2^o$ &  706.14166  &  707.17432  &   705.88123   & 706.04395  & 706.04382	& 707.36456   &  3.196-11     \\
   7  &   1s2p    &  $^1$P$_1^o$ &  707.66900  &  708.70074  &   707.42346   & 707.61078  & 707.61066	& 708.94946   &  1.123-15     \\
   8  &   1s3s    &  $^3$S$_1  $ &  830.87033  &  831.90155  &   830.60962   & 830.76263  & 830.76239	& 832.19769   &  1.617-13     \\
   9  &   1s3p    &  $^3$P$_0^o$ &  831.60281  &  832.58154  &   831.34540   & 831.50067  & 831.50061	& 832.83984   &  5.362-14     \\
  10  &   1s3s    &  $^1$S$_0  $ &  831.63510  &  832.63214  &   831.38556   & 831.53259  & 831.53241	& 832.90070   &  1.610-13     \\
  11  &   1s3p    &  $^3$P$_1^o$ &  831.65690  &  832.70105  &   831.40076   & 831.55780  & 831.55774	& 832.95496   &  1.565-14     \\
  12  &   1s3p    &  $^3$P$_2^o$ &  832.38076  &  833.43121  &   832.12354   & 832.27997  & 832.27979	& 833.66095   &  5.612-14     \\
  13  &   1s3d    &  $^3$D$_1  $ &  832.76987  &  833.80634  &   832.51001   & 832.65704  & 832.65704	& 834.03058   &  1.859-14     \\
  14  &   1s3d    &  $^3$D$_2  $ &  832.75629  &  833.81580  &   832.49683   & 832.64435  & 832.64435	& 834.04120   &  1.856-14     \\
  15  &   1s3p    &  $^1$P$_1^o$ &  832.79611  &  833.84375  &   832.54443   & 832.70276  & 832.70270	& 834.07385   &  3.797-15     \\
  16  &   1s3d    &  $^3$D$_3  $ &  833.01846  &  834.07513  &   832.75812   & 832.90497  & 832.90497	& 834.30786   &  1.892-14     \\
  17  &   1s3d    &  $^1$D$_2  $ &  833.04750  &  834.09094  &   832.78772   & 832.93549  & 832.93549	& 834.32520   &  1.891-14     \\
  18  &   1s4s    &  $^3$S$_1  $ &  875.91260  &  876.96307  &   875.65991   & 875.80420  & 875.80396	& 877.24005   &  2.391-13     \\
  19  &   1s4p    &  $^3$P$_0^o$ &  876.21880  &  877.24286  &   875.96198   & 876.09937  & 876.09924	& 877.50201   &  9.218-14     \\
  20  &   1s4s    &  $^1$S$_0  $ &  876.22970  &  877.26007  &   875.97601   & 876.11591  & 876.11560	& 877.51215   &  2.235-13     \\
  21  &   1s4p    &  $^3$P$_1^o$ &  876.24250  &  877.29211  &   875.98511   & 876.12311  & 876.12305	& 877.54706   &  3.362-14     \\
  22  &   1s4p    &  $^3$P$_2^o$ &  876.54690  &  877.60040  &   876.29034   & 876.42865  & 876.42859	& 877.83502   &  9.582-14     \\
  23  &   1s4d    &  $^3$D$_1  $ &  876.71090  &  877.75427  &   876.44928   & 876.59924  & 876.59924	& 877.98401   &  4.323-14     \\
  24  &   1s4d    &  $^3$D$_2  $ &  876.70540  &  877.75928  &   876.44476   & 876.59497  & 876.59503	& 877.98950   &  4.314-14     \\
  25  &   1s4p    &  $^1$P$_1^o$ &  876.72270  &  877.76935  &   876.46326   & 876.60138  & 876.60126	& 877.99951   &  8.917-15     \\
  26  &   1s4d    &  $^3$D$_3  $ &  876.81570  &  877.86780  &   876.55408   & 876.70355  & 876.70355	& 878.09973   &  4.408-14     \\
  27  &   1s4d    &  $^1$D$_2  $ &  876.82840  &  877.87640  &   876.56842   & 876.71838  & 876.71838	& 878.10712   &  4.397-14     \\
  28  &   1s4f    &  $^3$F$_2^o$ &  	       &  877.87708  &   876.56818   & 876.70862  & 876.70874	& 878.10901   &  8.863-14     \\
  29  &   1s4f    &  $^3$F$_3^o$ &  	       &  877.87720  &   876.56390   & 876.70441  & 876.70441	& 878.10699   &  8.861-14     \\
  30  &   1s4f    &  $^3$F$_4^o$ &  	       &  877.93347  &   876.62061   & 876.76111  & 876.76111	& 878.16522   &  8.912-14     \\
  31  &   1s4f    &  $^1$F$_3^o$ &  	       &  877.93359  &   876.62390   & 876.76440  & 876.76440	& 878.16541   &  8.917-14     \\
  32  &   1s5s    &  $^3$S$_1  $ &  896.64490  &  897.69702  &   896.38977   & 896.52985  & 896.52966	& 897.96375   &  3.879-13     \\
  33  &   1s5p    &  $^3$P$_0^o$ &  	       &  897.83850  &   896.54236   & 896.67773  & 896.67767	& 898.09497   &  1.594-13     \\
  34  &   1s5s    &  $^1$S$_0  $ &  896.80620  &  897.84711  &   896.54980   & 896.68646  & 896.68610	& 898.09277   &  3.163-13     \\
  35  &   1s5p    &  $^3$P$_1^o$ &  896.81250  &  897.86346  &   896.55420   & 896.68994  & 896.68976	& 898.11639   &  6.572-14     \\
  36  &   1s5p    &  $^3$P$_2^o$ &  896.96840  &  898.02118  &   896.71039   & 896.84631  & 896.84613	& 898.25702   &  1.655-13     \\
  37  &   1s5d    &  $^3$D$_1  $ &  	       &  898.09888  &   896.79077   & 896.93866  & 896.93866	& 898.33087   &  8.334-14     \\
  38  &   1s5d    &  $^3$D$_2  $ &  	       &  898.10168  &   896.78870   & 896.93671  & 896.93671	& 898.33380   &  8.250-14     \\
  39  &   1s5p    &  $^1$P$_1^o$ &  897.05860  &  898.10681  &   896.79816   & 896.93341  & 896.93323	& 898.33759   &  1.838-14     \\
  40  &   1s5d    &  $^3$D$_3  $ &  	       &  898.15704  &   896.84442   & 896.99194  & 896.99194	& 898.38898   &  8.523-14     \\
  41  &   1s5d    &  $^1$D$_2  $ &  	       &  898.16187  &   896.85217   & 897.00000  & 897.00000	& 898.39313   &  8.331-14     \\
  42  &   1s5f    &  $^3$F$_2^o$ &  	       &  898.16223  &   896.85205   & 896.99219  & 896.99219	& 898.39404   &  1.710-13     \\
  43  &   1s5f    &  $^3$F$_3^o$ &  	       &  898.16229  &   896.84991   & 896.99005  & 896.99005	& 898.39307   &  1.711-13     \\
  44  &   1s5f    &  $^3$F$_4^o$ &  	       &  898.19110  &   896.87891   & 897.01904  & 897.01904	& 898.42267   &  1.721-13     \\
  45  &   1s5f    &  $^1$F$_3^o$ &  	       &  898.19122  &   896.88068   & 897.02075  & 897.02075	& 898.42273   &  1.722-13     \\
  46  &   1s5g    &  $^3$G$_3  $ &  	       &  898.19122  &   896.88049   & 897.02026  & 897.02026	& 898.42194   &  2.886-13     \\
  47  &   1s5g    &  $^3$G$_4  $ &  	       &  898.19122  &   896.87921   & 897.01904  & 897.01904	& 898.42194   &  2.886-13     \\
  48  &   1s5g    &  $^3$G$_5  $ &  	       &  898.20850  &   896.89661   & 897.03638  & 897.03638	& 898.43976   &  2.893-13     \\
  49  &   1s5g    &  $^1$G$_4  $ &  	       &  898.20850  &   896.89758   & 897.03741  & 897.03741	& 898.43976   &  2.894-13     \\
\hline 													     
\end{tabular}      
			      
%\vspace*{0.5 cm}
\begin{flushleft}
{\small
NIST: {\tt http://nist.gov/pml/data/asd.cfm} \\
GRASP1: Energies from the {\sc grasp} code with 49 level calculations {\em without} Breit and QED effects \\
GRASP2: Energies from the {\sc grasp} code with 49 level calculations {\em with} Breit and QED effects \\
FAC1: Energies from the {\sc fac} code with 49 level calculations \\
FAC2: Energies from the {\sc fac} code with 71 level calculations \\
AS: Energies from the {\sc as} code with 49 level calculations

}
\end{flushleft}
\end{table*}

\setcounter{table}{0} 

\begin{table*} 
\caption{b. Energy levels (in Ryd) of Ge XXXI and their lifetimes ($\tau$, s). $a{\pm}b \equiv a{\times}$10$^{{\pm}b}$.} 
\begin{tabular}{rllrrrrrrr} \hline
Index  & \multicolumn{2}{c}{Configuration/Level} & NIST  & GRASP1 & GRASP2 &   FAC1  &    FAC2         &    AS        & $\tau$ (s)   \\    
\hline
   1  &   1s$^2$  &  $^1$S$_0  $ &    0.00000  &    0.00000  &    0.00000    &   0.00000  &    0.00000  &    0.00000  &  ........     \\
   2  &   1s2s    &  $^3$S$_1  $ &  748.31400  &  749.35278  &  747.98206    & 748.11768  &  748.11737  &  749.63153  &  5.872-10     \\
   3  &   1s2p    &  $^3$P$_0^o$ &  751.01363  &  751.90405  &  750.76447    & 750.93030  &  750.93018  &  752.07312  &  1.956-09     \\
   4  &   1s2s    &  $^1$S$_0  $ &  751.22441  &  752.21918  &  751.01996    & 751.16595  &  751.16565  &  752.57642  &  1.711-03     \\
   5  &   1s2p    &  $^3$P$_1^o$ &  751.28783  &  752.36719  &  750.96320    & 751.13751  &  751.13733  &  752.54663  &  5.186-15     \\
   6  &   1s2p    &  $^3$P$_2^o$ &  754.00687  &  755.18738  &  753.75824    & 753.92365  &  753.92346  &  755.36475  &  2.482-11     \\
   7  &   1s2p    &  $^1$P$_1^o$ &  755.57598  &  756.74915  &  755.34058    & 755.52960  &  755.52948  &  756.98779  &  1.001-15     \\
   8  &   1s3s    &  $^3$S$_1  $ &  887.14927  &  888.31311  &  886.88550    & 887.04102  &  887.04077  &  888.60968  &  1.413-13     \\
   9  &   1s3p    &  $^3$P$_0^o$ &  887.91273  &  889.01868  &  887.65234    & 887.81000  &  887.81000  &  889.27332  &  4.692-14     \\
  10  &   1s3s    &  $^1$S$_0  $ &  887.94618  &  889.07037  &  887.69275    & 887.84222  &  887.84198  &  889.34064  &  1.408-13     \\
  11  &   1s3p    &  $^3$P$_1^o$ &  887.96780  &  889.14655  &  887.70868    & 887.86829  &  887.86823  &  889.39624  &  1.307-14     \\
  12  &   1s3p    &  $^3$P$_2^o$ &  888.80359  &  889.98871  &  888.54340    & 888.70233  &  888.70221  &  890.20734  &  4.926-14     \\
  13  &   1s3d    &  $^3$D$_1  $ &  889.20682  &  890.37683  &  888.94409    & 889.09344  &  889.09344  &  890.58875  &  1.629-14     \\
  14  &   1s3d    &  $^3$D$_2  $ &  889.19124  &  890.38666  &  888.92889    & 889.07880  &  889.07880  &  890.59973  &  1.626-14     \\
  15  &   1s3p    &  $^1$P$_1^o$ &  889.23052  &  890.41095  &  888.97571    & 889.13641  &  889.13635  &  890.62976  &  3.384-15     \\
  16  &   1s3d    &  $^3$D$_3  $ &  889.49123  &  890.68347  &  889.22791    & 889.37708  &  889.37708  &  890.90466  &  1.660-14     \\
  17  &   1s3d    &  $^1$D$_2  $ &  889.52231  &  890.69977  &  889.25946    & 889.40955  &  889.40955  &  890.92255  &  1.658-14     \\
  18  &   1s4s    &  $^3$S$_1  $ &  	       &  936.45941  &  935.01917    & 935.16589  &  935.16571  &  936.73291  &  2.088-13     \\
  19  &   1s4p    &  $^3$P$_0^o$ &  	       &  936.74963  &  935.33398    & 935.47363  &  935.47357  &  937.00336  &  8.067-14     \\
  20  &   1s4s    &  $^1$S$_0  $ &  	       &  936.76715  &  935.34802    & 935.49023  &  935.48993  &  937.01495  &  1.958-13     \\
  21  &   1s4p    &  $^3$P$_1^o$ &  	       &  936.80231  &  935.35748    & 935.49786  &  935.49774  &  937.05133  &  2.829-14     \\
  22  &   1s4p    &  $^3$P$_2^o$ &  	       &  937.15784  &  935.70990    & 935.85071  &  935.85052  &  937.38153  &  8.407-14     \\
  23  &   1s4d    &  $^3$D$_1  $ &  	       &  937.31708  &  935.87476    & 936.02710  &  936.02710  &  937.53510  &  3.788-14     \\
  24  &   1s4d    &  $^3$D$_2  $ &  	       &  937.32227  &  935.86945    & 936.02203  &  936.02203  &  937.54077  &  3.778-14     \\
  25  &   1s4p    &  $^1$P$_1^o$ &  	       &  937.33075  &  935.88757    & 936.02808  &  936.02795  &  937.54968  &  7.964-15     \\
  26  &   1s4d    &  $^3$D$_3  $ &  	       &  937.44659  &  935.99463    & 936.14648  &  936.14648  &  937.66681  &  3.868-14     \\
  27  &   1s4d    &  $^1$D$_2  $ &  	       &  937.45551  &  936.00989    & 936.16217  &  936.16217  &  937.67432  &  3.854-14     \\
  28  &   1s4f    &  $^3$F$_2^o$ &  	       &  937.45624  &  936.00952    & 936.15247  &  936.15247  &  937.67645  &  7.768-14     \\
  29  &   1s4f    &  $^3$F$_3^o$ &  	       &  937.45630  &  936.00482    & 936.14771  &  936.14771  &  937.67419  &  7.767-14     \\
  30  &   1s4f    &  $^3$F$_4^o$ &  	       &  937.52051  &  936.06946    & 936.21234  &  936.21234  &  937.74054  &  7.814-14     \\
  31  &   1s4f    &  $^1$F$_3^o$ &  	       &  937.52063  &  936.07312    & 936.21600  &  936.21600  &  937.74066  &  7.818-14     \\
  32  &   1s5s    &  $^3$S$_1  $ &  	       &  958.61139  &  957.16663    & 957.30908  &  957.30884  &  958.87256  &  3.383-13     \\
  33  &   1s5p    &  $^3$P$_0^o$ &  	       &  958.75806  &  957.32556    & 957.46332  &  957.46326  &  959.00800  &  1.394-13     \\
  34  &   1s5s    &  $^1$S$_0  $ &  	       &  958.76678  &  957.33301    & 957.47198  &  957.47162  &  959.00586  &  2.783-13     \\
  35  &   1s5p    &  $^3$P$_1^o$ &  	       &  958.78479  &  957.33759    & 957.47565  &  957.47552  &  959.03070  &  5.538-14     \\
  36  &   1s5p    &  $^3$P$_2^o$ &  	       &  958.96661  &  957.51794    & 957.65625  &  957.65607  &  959.19153  &  1.451-13     \\
  37  &   1s5d    &  $^3$D$_1  $ &  	       &  959.04706  &  957.60132    & 957.75153  &  957.75153  &  959.26764  &  7.301-14     \\
  38  &   1s5d    &  $^3$D$_2  $ &  	       &  959.04999  &  957.59882    & 957.74921  &  957.74921  &  959.27075  &  7.220-14     \\
  39  &   1s5p    &  $^1$P$_1^o$ &  	       &  959.05426  &  957.60809    & 957.74567  &  957.74554  &  959.27380  &  1.637-14     \\
  40  &   1s5d    &  $^3$D$_3  $ &  	       &  959.11340  &  957.66266    & 957.81256  &  957.81256  &  959.33368  &  7.477-14     \\
  41  &   1s5d    &  $^1$D$_2  $ &  	       &  959.11841  &  957.67090    & 957.82111  &  957.82111  &  959.33795  &  7.299-14     \\
  42  &   1s5f    &  $^3$F$_2^o$ &  	       &  959.11877  &  957.67078    & 957.81323  &  957.81323  &  959.33893  &  1.499-13     \\
  43  &   1s5f    &  $^3$F$_3^o$ &  	       &  959.11884  &  957.66833    & 957.81085  &  957.81085  &  959.33777  &  1.500-13     \\
  44  &   1s5f    &  $^3$F$_4^o$ &  	       &  959.15173  &  957.70148    & 957.84387  &  957.84387  &  959.37146  &  1.509-13     \\
  45  &   1s5f    &  $^1$F$_3^o$ &  	       &  959.15179  &  957.70337    & 957.84583  &  957.84583  &  959.37152  &  1.510-13     \\
  46  &   1s5g    &  $^3$G$_3  $ &  	       &  959.15186  &  957.70312    & 957.84528  &  957.84528  &  959.37073  &  2.531-13     \\
  47  &   1s5g    &  $^3$G$_4  $ &  	       &  959.15186  &  957.70172    & 957.84393  &  957.84393  &  959.37073  &  2.530-13     \\
  48  &   1s5g    &  $^3$G$_5  $ &  	       &  959.17157  &  957.72156    & 957.86377  &  957.86377  &  959.39099  &  2.537-13     \\
  49  &   1s5g    &  $^1$G$_4  $ &  	       &  959.17157  &  957.72272    & 957.86487  &  957.86487  &  959.39099  &  2.538-13     \\
\hline 													     
\end{tabular}      
			      
%\vspace*{0.5 cm}
\begin{flushleft}
{\small
NIST: {\tt http://nist.gov/pml/data/asd.cfm} \\
GRASP1: Energies from the {\sc grasp} code with 49 level calculations {\em without} Breit and QED effects \\
GRASP2: Energies from the {\sc grasp} code with 49 level calculations {\em with} Breit and QED effects \\
FAC1: Energies from the {\sc fac} code with 49 level calculations \\
FAC2: Energies from the {\sc fac} code with 71 level calculations \\
AS: Energies from the {\sc as} code with 49 level calculations

}
\end{flushleft}
\end{table*}

\setcounter{table}{0} 

\begin{table*} 
\caption{c. Energy levels (in Ryd) of As XXXII and their lifetimes ($\tau$, s). $a{\pm}b \equiv a{\times}$10$^{{\pm}b}$.} 
\begin{tabular}{rllrrrrrrr} \hline
Index  & \multicolumn{2}{c}{Configuration/Level} & NIST  & GRASP1 & GRASP2 &   FAC1  &    FAC2         &    AS        & $\tau$ (s)   \\    
\hline
   1  &   1s$^2$  &  $^1$S$_0  $ &     0.00000  &     0.00000  &      0.00000	&    0.00000  &     0.00000  &    0.00000   &  ........     \\
   2  &   1s2s    &  $^3$S$_1  $ &   797.34582  &   798.48352  &    796.97406	&  797.11176  &   797.11145  &  798.76208   &  4.281-10     \\
   3  &   1s2p    &  $^3$P$_0^o$ &   800.26160  &   801.12836  &    799.87250	&  800.04065  &   800.04053  &  801.28674   &  1.848-09     \\
   4  &   1s2s    &  $^1$S$_0  $ &   800.43502  &   801.45300  &    800.13220	&  800.27991  &   800.27954  &  801.83307   &  1.345-03     \\
   5  &   1s2p    &  $^3$P$_1^o$ &   800.46973  &   801.62323  &    800.07312	&  800.25043  &   800.25031  &  801.79297   &  4.269-15     \\
   6  &   1s2p    &  $^3$P$_2^o$ &   803.56011  &   804.86084  &    803.28619	&  803.45416  &   803.45398  &  805.02167   &  1.941-11     \\
   7  &   1s2p    &  $^1$P$_1^o$ &   805.17105  &   806.45795  &    804.90912	&  805.09967  &   805.09955  &  806.68280   &  8.956-16     \\
   8  &   1s3s    &  $^3$S$_1  $ &   945.39582  &   946.65387  &    945.08112	&  945.23901  &   945.23883  &  946.94891   &  1.239-13     \\
   9  &   1s3p    &  $^3$P$_0^o$ &   946.19956  &   947.38519  &    945.87970	&  946.03967  &   946.03955  &  947.63403   &  4.122-14     \\
  10  &   1s3s    &  $^1$S$_0  $ &   946.21232  &   947.43805  &    945.92023	&  946.07190  &   946.07166  &  947.70905   &  1.236-13     \\
  11  &   1s3p    &  $^3$P$_1^o$ &   946.25059  &   947.52148  &    945.93640	&  946.09845  &   946.09827  &  947.76471   &  1.100-14     \\
  12  &   1s3p    &  $^3$P$_2^o$ &   947.19649  &   948.48816  &    946.89569	&  947.05701  &   947.05695  &  948.69220   &  4.343-14     \\
  13  &   1s3d    &  $^3$D$_1  $ &  	        &   948.88940  &    947.31073	&  947.46228  &   947.46228  &  949.08521   &  1.434-14     \\
  14  &   1s3d    &  $^3$D$_2  $ &  	        &   948.89960  &    947.29333	&  947.44543  &   947.44543  &  949.09668   &  1.430-14     \\
  15  &   1s3p    &  $^1$P$_1^o$ &   947.62570  &   948.92029  &    947.33972	&  947.50256  &   947.50244  &  949.12433   &  3.026-15     \\
  16  &   1s3d    &  $^3$D$_3  $ &   948.02483  &   949.23773  &    947.63397	&  947.78534  &   947.78534  &  949.44379   &  1.463-14     \\
  17  &   1s3d    &  $^1$D$_2  $ &   948.05855  &   949.25452  &    947.66754	&  947.81982  &   947.81982  &  949.46222   &  1.460-14     \\
  18  &   1s4s    &  $^3$S$_1  $ &   996.69743  &   997.99231  &    996.40552	&  996.55450  &   996.55432  &  998.25989   &  1.830-13     \\
  19  &   1s4p    &  $^3$P$_0^o$ &   997.93277  &   998.29315  &    996.73328	&  996.87506  &   996.87506  &  998.53888   &  7.088-14     \\
  20  &   1s4s    &  $^1$S$_0  $ &   997.02822  &   998.31085  &    996.74725	&  996.89166  &   996.89142  &  998.55237   &  1.722-13     \\
  21  &   1s4p    &  $^3$P$_1^o$ &   997.05373  &   998.34918  &    996.75690	&  996.89948  &   996.89941  &  998.58966   &  2.400-14     \\
  22  &   1s4p    &  $^3$P$_2^o$ &   997.44649  &   998.75720  &    997.16187	&  997.30493  &   997.30481  &  998.96649   &  7.407-14     \\
  23  &   1s4d    &  $^3$D$_1  $ &  	        &   998.92188  &    997.33270	&  997.48724  &   997.48724  &  999.12457   &  3.333-14     \\
  24  &   1s4d    &  $^3$D$_2  $ &  	        &   998.92731  &    997.32654	&  997.48126  &   997.48138  &  999.13049   &  3.322-14     \\
  25  &   1s4p    &  $^1$P$_1^o$ &  997.62874   &   998.93420  &    997.34442	&  997.48712  &   997.48694  &  999.13831   &  7.137-15     \\
  26  &   1s4d    &  $^3$D$_3  $ &  	        &   999.06903  &    997.46918	&  997.62323  &   997.62323  &  999.27405   &  3.408-14     \\
  27  &   1s4d    &  $^1$D$_2  $ &  	        &   999.07819  &    997.48535	&  997.63983  &   997.63983  &  999.28149   &  3.391-14     \\
  28  &   1s4f    &  $^3$F$_2^o$ &  	        &   999.07898  &    997.48499	&  997.63007  &   997.63007  &  999.28387   &  6.838-14     \\
  29  &   1s4f    &  $^3$F$_3^o$ &  	        &   999.07904  &    997.47980	&  997.62482  &   997.62482  &  999.28143   &  6.836-14     \\
  30  &   1s4f    &  $^3$F$_4^o$ &  	        &   999.15198  &    997.55322	&  997.69824  &   997.69824  &  999.35663   &  6.881-14     \\
  31  &   1s4f    &  $^1$F$_3^o$ &  	        &   999.15210  &    997.55725	&  997.70227  &   997.70227  &  999.35681   &  6.885-14     \\
  32  &   1s5s    &  $^3$S$_1  $ &  1020.30744  &  1021.61121  &   1020.01941	& 1020.16406  &  1020.16388  & 1021.86420   &  2.962-13     \\
  33  &   1s5p    &  $^3$P$_0^o$ &  1020.47694  &  1021.76324  &   1020.18488	& 1020.32477  &  1020.32471  & 1022.00391   &  1.225-13     \\
  34  &   1s5s    &  $^1$S$_0  $ &  1020.47147  &  1021.77209  &   1020.19226	& 1020.33350  &  1020.33301  & 1022.00189   &  2.458-13     \\
  35  &   1s5p    &  $^3$P$_1^o$ &  1020.48696  &  1021.79169  &   1020.19696	& 1020.33722  &  1020.33716  & 1022.02777   &  4.702-14     \\
  36  &   1s5p    &  $^3$P$_2^o$ &  1020.68835  &  1022.00037  &   1020.40417	& 1020.54462  &  1020.54449  & 1022.21088   &  1.277-13     \\
  37  &   1s5d    &  $^3$D$_1  $ &  	        &  1022.08356  &   1020.49054	& 1020.64294  &  1020.64294  & 1022.28912   &  6.421-14     \\
  38  &   1s5d    &  $^3$D$_2  $ &  	        &  1022.08655  &   1020.48761	& 1020.64026  &  1020.64026  & 1022.29230   &  6.345-14     \\
  39  &   1s5p    &  $^1$P$_1^o$ &  1020.78039  &  1022.09009  &   1020.49677	& 1020.63660  &  1020.63641  & 1022.29474   &  1.463-14     \\
  40  &   1s5d    &  $^3$D$_3  $ &  	        &  1022.15894  &   1020.56042	& 1020.71252  &  1020.71252  & 1022.36395   &  6.586-14     \\
  41  &   1s5d    &  $^1$D$_2  $ &  	        &  1022.16406  &   1020.56915	& 1020.72150  &  1020.72150  & 1022.36816   &  6.421-14     \\
  42  &   1s5f    &  $^3$F$_2^o$ &  	        &  1022.16449  &   1020.56897	& 1020.71362  &  1020.71362  & 1022.36932   &  1.319-13     \\
  43  &   1s5f    &  $^3$F$_3^o$ &  	        &  1022.16455  &   1020.56635	& 1020.71100  &  1020.71100  & 1022.36810   &  1.320-13     \\
  44  &   1s5f    &  $^3$F$_4^o$ &  	        &  1022.20190  &   1020.60394	& 1020.74854  &  1020.74854  & 1022.40625   &  1.329-13     \\
  45  &   1s5f    &  $^1$F$_3^o$ &  	        &  1022.20197  &   1020.60602	& 1020.75061  &  1020.75061  & 1022.40637   &  1.329-13     \\
  46  &   1s5g    &  $^3$G$_3  $ &  	        &  1022.20203  &   1020.60577	& 1020.75012  &  1020.75012  & 1022.40533   &  2.228-13     \\
  47  &   1s5g    &  $^3$G$_4  $ &  	        &  1022.20203  &   1020.60425	& 1020.74854  &  1020.74860  & 1022.40533   &  2.228-13     \\
  48  &   1s5g    &  $^3$G$_5  $ &  	        &  1022.22437  &   1020.62677	& 1020.77106  &  1020.77106  & 1022.42841   &  2.234-13     \\
  49  &   1s5g    &  $^1$G$_4  $ &  	        &  1022.22437  &   1020.62799	& 1020.77228  &  1020.77228  & 1022.42841   &  2.235-13     \\
\hline 													     
\end{tabular}      
			      
%\vspace*{0.5 cm}
\begin{flushleft}
{\small
NIST: {\tt http://nist.gov/pml/data/asd.cfm} \\
GRASP1: Energies from the {\sc grasp} code with 49 level calculations {\em without} Breit and QED effects \\
GRASP2: Energies from the {\sc grasp} code with 49 level calculations {\em with} Breit and QED effects \\
FAC1: Energies from the {\sc fac} code with 49 level calculations \\
FAC2: Energies from the {\sc fac} code with 71 level calculations \\
AS: Energies from the {\sc as} code with 49 level calculations

}
\end{flushleft}
\end{table*}

\setcounter{table}{0} 

\begin{table*} 
\caption{d. Energy levels (in Ryd) of Se XXXIII and their lifetimes ($\tau$, s). $a{\pm}b \equiv a{\times}$10$^{{\pm}b}$.} 
\begin{tabular}{rllrrrrrrr} \hline
Index  & \multicolumn{2}{c}{Configuration/Level} & NIST  & GRASP1 & GRASP2 &   FAC1  &    FAC2         &    AS        & $\tau$ (s)   \\    
\hline
   1  &   1s$^2$  &  $^1$S$_0  $ &     0.00000  &     0.00000  &     0.00000	&    0.00000  &     0.00000  &     0.00000  &  ........     \\
   2  &   1s2s    &  $^3$S$_1  $ &   847.96060  &   849.23895  &   847.58148	&  847.72168  &   847.72144  &   849.51550  &  3.150-10     \\
   3  &   1s2p    &  $^3$P$_0^o$ &   850.99949  &   851.97845  &   850.59851	&  850.76965  &   850.76959  &   852.12384  &  1.747-09     \\
   4  &   1s2s    &  $^1$S$_0  $ &   851.16899  &   852.31256  &   850.86206	&  851.01196  &   851.01160  &   852.71826  &  1.064-03     \\
   5  &   1s2p    &  $^3$P$_1^o$ &   851.20918  &   852.50494  &   850.79877	&  850.97968  &   850.97961  &   852.66272  &  3.554-15     \\
   6  &   1s2p    &  $^3$P$_2^o$ &   854.73715  &   856.20563  &   854.47565	&  854.64679  &   854.64667  &   856.34576  &  1.527-11     \\
   7  &   1s2p    &  $^1$P$_1^o$ &   856.38873  &   857.83801  &   856.13965	&  856.33240  &   856.33221  &   858.04529  &  8.040-16     \\
   8  &   1s3s    &  $^3$S$_1  $ &  1005.51485  &  1006.93396  &  1005.20636	& 1005.36725  &  1005.36713  &  1007.22571  &  1.091-13     \\
   9  &   1s3p    &  $^3$P$_0^o$ &  1006.35230  &  1007.69147  &  1006.03735	& 1006.20013  &  1006.20013  &  1007.93231  &  3.636-14     \\
  10  &   1s3s    &  $^1$S$_0  $ &  1006.36232  &  1007.74530  &  1006.07794	& 1006.23242  &  1006.23224  &  1008.01617  &  1.089-13     \\
  11  &   1s3p    &  $^3$P$_1^o$ &  1006.40242  &  1007.83606  &  1006.09381	& 1006.25885  &  1006.25879  &  1008.07056  &  9.342-15     \\
  12  &   1s3p    &  $^3$P$_2^o$ &  1007.46131  &  1008.94080  &  1007.19128	& 1007.35559  &  1007.35553  &  1009.12634  &  3.843-14     \\
  13  &   1s3d    &  $^3$D$_1  $ &  1007.99431  &  1009.35510  &  1007.62073	& 1007.77502  &  1007.77502  &  1009.53088  &  1.267-14     \\
  14  &   1s3d    &  $^3$D$_2  $ &  	        &  1009.36566  &  1007.60101	& 1007.75592  &  1007.75592  &  1009.54279  &  1.262-14     \\
  15  &   1s3p    &  $^1$P$_1^o$ &  1007.92059  &  1009.38281  &  1007.64716	& 1007.81281  &  1007.81274  &  1009.56805  &  2.715-15     \\
  16  &   1s3d    &  $^3$D$_3  $ &  1008.35800  &  1009.74933  &  1007.98737	& 1008.14160  &  1008.14160  &  1009.93610  &  1.294-14     \\
  17  &   1s3d    &  $^1$D$_2  $ &  1008.39354  &  1009.76660  &  1008.02307	& 1008.17822  &  1008.17822  &  1009.95514  &  1.290-14     \\
  18  &   1s4s    &  $^3$S$_1  $ &  1060.11159  &  1061.57275  &  1059.82959	& 1059.98145  &  1059.98120  &  1061.83191  &  1.611-13     \\
  19  &   1s4p    &  $^3$P$_0^o$ &  1060.46060  &  1061.88428  &  1060.17053	& 1060.31519  &  1060.31506  &  1062.11938  &  6.252-14     \\
  20  &   1s4s    &  $^1$S$_0  $ &  1060.45422  &  1061.90222  &  1060.18445	& 1060.33167  &  1060.33142  &  1062.13513  &  1.519-13     \\
  21  &   1s4p    &  $^3$P$_1^o$ &  1060.48065  &  1061.94373  &  1060.19397	& 1060.33948  &  1060.33936  &  1062.17285  &  2.051-14     \\
  22  &   1s4p    &  $^3$P$_2^o$ &  1060.92808  &  1062.40991  &  1060.65723	& 1060.80322  &  1060.80310  &  1062.60095  &  6.552-14     \\
  23  &   1s4d    &  $^3$D$_1  $ &  	        &  1062.57996  &  1060.83411	& 1060.99146  &  1060.99146  &  1062.76343  &  2.944-14     \\
  24  &   1s4d    &  $^3$D$_2  $ &  	        &  1062.58557  &  1060.82690	& 1060.98462  &  1060.98462  &  1062.76965  &  2.932-14     \\
  25  &   1s4p    &  $^1$P$_1^o$ &  	        &  1062.59106  &  1060.84473	& 1060.99023  &  1060.99011  &  1062.77637  &  6.417-15     \\
  26  &   1s4d    &  $^3$D$_3  $ &  	        &  1062.74658  &  1060.98889	& 1061.14575  &  1061.14575  &  1062.93225  &  3.014-14     \\
  27  &   1s4d    &  $^1$D$_2  $ &  	        &  1062.75598  &  1061.00598	& 1061.16333  &  1061.16333  &  1062.93970  &  2.996-14     \\
  28  &   1s4f    &  $^3$F$_2^o$ &  	        &  1062.75684  &  1061.00562	& 1061.15344  &  1061.15344  &  1062.94238  &  6.042-14     \\
  29  &   1s4f    &  $^3$F$_3^o$ &  	        &  1062.75684  &  1060.99988	& 1061.14771  &  1061.14771  &  1062.93958  &  6.040-14     \\
  30  &   1s4f    &  $^3$F$_4^o$ &  	        &  1062.83948  &  1061.08289	& 1061.23083  &  1061.23083  &  1063.02466  &  6.082-14     \\
  31  &   1s4f    &  $^1$F$_3^o$ &  	        &  1062.83960  &  1061.08728	& 1061.23511  &  1061.23523  &  1063.02478  &  6.086-14     \\
  32  &   1s5s    &  $^3$S$_1  $ &  1085.23431  &  1086.70789  &  1084.95898	& 1085.10657  &  1085.10632  &  1086.94958  &  2.603-13     \\
  33  &   1s5p    &  $^3$P$_0^o$ &  1085.41110  &  1086.86523  &  1085.13110	& 1085.27393  &  1085.27380  &  1087.09338  &  1.080-13     \\
  34  &   1s5s    &  $^1$S$_0  $ &  1085.40563  &  1086.87415  &  1085.13855	& 1085.28247  &  1085.28210  &  1087.09192  &  2.177-13     \\
  35  &   1s5p    &  $^3$P$_1^o$ &  1085.42203  &  1086.89539  &  1085.14319	& 1085.28625  &  1085.28613  &  1087.11865  &  4.020-14     \\
  36  &   1s5p    &  $^3$P$_2^o$ &  1085.74553  &  1087.13391  &  1085.38013	& 1085.52356  &  1085.52332  &  1087.32605  &  1.129-13     \\
  37  &   1s5d    &  $^3$D$_1  $ &  	        &  1087.21973  &  1085.46960	& 1085.62488  &  1085.62488  &  1087.40637  &  5.670-14     \\
  38  &   1s5d    &  $^3$D$_2  $ &  	        &  1087.22290  &  1085.46619	& 1085.62158  &  1085.62158  &  1087.40967  &  5.597-14     \\
  39  &   1s5p    &  $^1$P$_1^o$ &  	        &  1087.22559  &  1085.47522	& 1085.61792  &  1085.61780  &  1087.41150  &  1.311-14     \\
  40  &   1s5d    &  $^3$D$_3  $ &  	        &  1087.30505  &  1085.54883	& 1085.70386  &  1085.70386  &  1087.49072  &  5.824-14     \\
  41  &   1s5d    &  $^1$D$_2  $ &  	        &  1087.31030  &  1085.55811	& 1085.71326  &  1085.71326  &  1087.49512  &  5.671-14     \\
  42  &   1s5f    &  $^3$F$_2^o$ &  	        &  1087.31079  &  1085.55786	& 1085.70544  &  1085.70544  &  1087.49634  &  1.165-13     \\
  43  &   1s5f    &  $^3$F$_3^o$ &  	        &  1087.31091  &  1085.55505	& 1085.70239  &  1085.70239  &  1087.49487  &  1.166-13     \\
  44  &   1s5f    &  $^3$F$_4^o$ &  	        &  1087.35315  &  1085.59753	& 1085.74487  &  1085.74487  &  1087.53796  &  1.174-13     \\
  45  &   1s5f    &  $^1$F$_3^o$ &  	        &  1087.35315  &  1085.59985	& 1085.74719  &  1085.74719  &  1087.53809  &  1.175-13     \\
  46  &   1s5g    &  $^3$G$_3  $ &  	        &  1087.35327  &  1085.59961	& 1085.74670  &  1085.74670  &  1087.53711  &  1.969-13     \\
  47  &   1s5g    &  $^3$G$_4  $ &  	        &  1087.35327  &  1085.59790	& 1085.74500  &  1085.74500  &  1087.53711  &  1.969-13     \\
  48  &   1s5g    &  $^3$G$_5  $ &  	        &  1087.37854  &  1085.62329	& 1085.77051  &  1085.77051  &  1087.56311  &  1.975-13     \\
  49  &   1s5g    &  $^1$G$_4  $ &  	        &  1087.37854  &  1085.62463	& 1085.77185  &  1085.77185  &  1087.56311  &  1.975-13     \\
\hline 													     
\end{tabular}      
			      
%\vspace*{0.5 cm}
\begin{flushleft}
{\small
NIST: {\tt http://nist.gov/pml/data/asd.cfm} \\
GRASP1: Energies from the {\sc grasp} code with 49 level calculations {\em without} Breit and QED effects \\
GRASP2: Energies from the {\sc grasp} code with 49 level calculations {\em with} Breit and QED effects \\
FAC1: Energies from the {\sc fac} code with 49 level calculations \\
FAC2: Energies from the {\sc fac} code with 71 level calculations \\
AS: Energies from the {\sc as} code with 49 level calculations

}
\end{flushleft}
\end{table*}

\setcounter{table}{0} 

\begin{table*} 
\caption{e. Energy levels (in Ryd) of Br XXXIV and their lifetimes ($\tau$, s). $a{\pm}b \equiv a{\times}$10$^{{\pm}b}$.} 
\begin{tabular}{rllrrrrrrr} \hline
Index  & \multicolumn{2}{c}{Configuration/Level} & NIST  & GRASP1 & GRASP2 &   FAC1  &    FAC2         &    AS        & $\tau$ (s)   \\    
\hline
   1  &   1s$^2$  &  $^1$S$_0  $ &     0.00000  &     0.00000  &     0.00000	&    0.00000  &     0.00000  &     0.00000  &  ........     \\
   2  &   1s2s    &  $^3$S$_1  $ &   900.19625  &   901.62781  &   899.81268	&  899.95477  &   899.95453  &   901.89984  &  2.338-10     \\
   3  &   1s2p    &  $^3$P$_0^o$ &   903.35862  &   904.46326  &   902.95123	&  903.12457  &   903.12451  &   904.59241  &  1.652-09     \\
   4  &   1s2s    &  $^1$S$_0  $ &   903.52584  &   904.80664  &   903.21796	&  903.36945  &   903.36914  &   905.24097  &  8.463-04     \\
   5  &   1s2p    &  $^3$P$_1^o$ &   903.57058  &   905.02130  &   903.14862	&  903.33252  &   903.33234  &   905.16406  &  2.988-15     \\
   6  &   1s2p    &  $^3$P$_2^o$ &   907.58070  &   909.23376  &   907.33832	&  907.51196  &   907.51178  &   909.34802  &  1.209-11     \\
   7  &   1s2p    &  $^1$P$_1^o$ &   909.27329  &   910.90149  &   909.04401	&  909.23810  &   909.23798  &   911.08606  &  7.240-16     \\
   8  &   1s3s    &  $^3$S$_1  $ &  1067.56849  &  1069.16479  &  1067.27222	& 1067.43542  &  1067.43518  &  1069.45044  &  9.630-14     \\
   9  &   1s3p    &  $^3$P$_0^o$ &  1068.44057  &  1069.94873  &  1068.13647	& 1068.30127  &  1068.30127  &  1070.17847  &  3.218-14     \\
  10  &   1s3s    &  $^1$S$_0  $ &  1068.44786  &  1070.00366  &  1068.17688	& 1068.33350  &  1068.33325  &  1070.27258  &  9.623-14     \\
  11  &   1s3p    &  $^3$P$_1^o$ &  1068.48978  &  1070.10168  &  1068.19202	& 1068.35925  &  1068.35913  &  1070.32410  &  7.994-15     \\
  12  &   1s3p    &  $^3$P$_2^o$ &  1069.69265  &  1071.35901  &  1069.44214	& 1069.60864  &  1069.60864  &  1071.52100  &  3.414-14     \\
  13  &   1s3d    &  $^3$D$_1  $ &  	        &  1071.78638  &  1069.88599	& 1070.04248  &  1070.04248  &  1071.93677  &  1.123-14     \\
  14  &   1s3d    &  $^3$D$_2  $ &  	        &  1071.79724  &  1069.86377	& 1070.02075  &  1070.02075  &  1071.94910  &  1.118-14     \\
  15  &   1s3p    &  $^1$P$_1^o$ &  1070.16469  &  1071.81091  &  1069.91003	& 1070.07776  &  1070.07764  &  1071.97241  &  2.444-15     \\
  16  &   1s3d    &  $^3$D$_3  $ &  	        &  1072.23083  &  1070.30029	& 1070.45667  &  1070.45667  &  1072.39319  &  1.149-14     \\
  17  &   1s3d    &  $^1$D$_2  $ &  	        &  1072.24866  &  1070.33826	& 1070.49548  &  1070.49548  &  1072.41284  &  1.144-14     \\
  18  &   1s4s    &  $^3$S$_1  $ &  1125.56790  &  1127.21277  &  1125.30298	& 1125.45703  &  1125.45691  &  1127.45984  &  1.422-13     \\
  19  &   1s4p    &  $^3$P$_0^o$ &  1125.93150  &  1127.53516  &  1125.65747	& 1125.80408  &  1125.80408  &  1127.75574  &  5.534-14     \\
  20  &   1s4s    &  $^1$S$_0  $ &  1125.92329  &  1127.55334  &  1125.67126	& 1125.82056  &  1125.82031  &  1127.77429  &  1.345-13     \\
  21  &   1s4p    &  $^3$P$_1^o$ &  1125.95063  &  1127.59790  &  1125.68066	& 1125.82812  &  1125.82812  &  1127.81189  &  1.766-14     \\
  22  &   1s4p    &  $^3$P$_2^o$ &  1126.45912  &  1128.12854  &  1126.20813	& 1126.35620  &  1126.35620  &  1128.29602  &  5.817-14     \\
  23  &   1s4d    &  $^3$D$_1  $ &  	        &  1128.30396  &  1126.39099	& 1126.55042  &  1126.55042  &  1128.46301  &  2.609-14     \\
  24  &   1s4d    &  $^3$D$_2  $ &  	        &  1128.30981  &  1126.38281	& 1126.54260  &  1126.54260  &  1128.46948  &  2.597-14     \\
  25  &   1s4p    &  $^1$P$_1^o$ &  1126.65231  &  1128.31372  &  1126.40063	& 1126.54822  &  1126.54810  &  1128.47510  &  5.785-15     \\
  26  &   1s4d    &  $^3$D$_3  $ &  	        &  1128.49182  &  1126.56604	& 1126.72498  &  1126.72498  &  1128.65295  &  2.676-14     \\
  27  &   1s4d    &  $^1$D$_2  $ &  	        &  1128.50146  &  1126.58411	& 1126.74341  &  1126.74341  &  1128.66040  &  2.657-14     \\
  28  &   1s4f    &  $^3$F$_2^o$ &  	        &  1128.50232  &  1126.58362	& 1126.73352  &  1126.73352  &  1128.66345  &  5.358-14     \\
  29  &   1s4f    &  $^3$F$_3^o$ &  	        &  1128.50244  &  1126.57739	& 1126.72729  &  1126.72729  &  1128.66028  &  5.357-14     \\
  30  &   1s4f    &  $^3$F$_4^o$ &  	        &  1128.59546  &  1126.67102	& 1126.82092  &  1126.82092  &  1128.75610  &  5.396-14     \\
  31  &   1s4f    &  $^1$F$_3^o$ &  	        &  1128.59558  &  1126.67578	& 1126.82568  &  1126.82568  &  1128.75623  &  5.399-14     \\
  32  &   1s5s    &  $^3$S$_1  $ &  1152.25436  &  1153.91357  &  1151.99756	& 1152.14722  &  1152.14697  &  1154.14014  &  2.296-13     \\
  33  &   1s5p    &  $^3$P$_0^o$ &  1152.43752  &  1154.07654  &  1152.17651	& 1152.32117  &  1152.32117  &  1154.28821  &  9.553-14     \\
  34  &   1s5s    &  $^1$S$_0  $ &  1152.43205  &  1154.08545  &  1152.18384	& 1152.32983  &  1152.32947  &  1154.28723  &  1.935-13     \\
  35  &   1s5p    &  $^3$P$_1^o$ &  1152.44846  &  1154.10828  &  1152.18835	& 1152.33350  &  1152.33337  &  1154.31445  &  3.460-14     \\
  36  &   1s5p    &  $^3$P$_2^o$ &  1152.70908  &  1154.37964  &  1152.45825	& 1152.60364  &  1152.60352  &  1154.54846  &  1.002-13     \\
  37  &   1s5d    &  $^3$D$_1  $ &  	        &  1154.46826  &  1152.55066	& 1152.70801  &  1152.70801  &  1154.63086  &  5.024-14     \\
  38  &   1s5d    &  $^3$D$_2  $ &  	        &  1154.47156  &  1152.54675	& 1152.70435  &  1152.70435  &  1154.63428  &  4.956-14     \\
  39  &   1s5p    &  $^1$P$_1^o$ &  1152.80658  &  1154.47351  &  1152.55591	& 1152.70056  &  1152.70032  &  1154.63550  &  1.178-14     \\
  40  &   1s5d    &  $^3$D$_3  $ &  	        &  1154.56445  &  1152.64026	& 1152.79724  &  1152.79724  &  1154.72571  &  5.169-14     \\
  41  &   1s5d    &  $^1$D$_2  $ &  	        &  1154.56995  &  1152.65002	& 1152.80725  &  1152.80725  &  1154.72998  &  5.028-14     \\
  42  &   1s5f    &  $^3$F$_2^o$ &  	        &  1154.57043  &  1152.64978	& 1152.79932  &  1152.79932  &  1154.73145  &  1.033-13     \\
  43  &   1s5f    &  $^3$F$_3^o$ &  	        &  1154.57043  &  1152.64661	& 1152.79614  &  1152.79614  &  1154.72986  &  1.034-13     \\
  44  &   1s5f    &  $^3$F$_4^o$ &  	        &  1154.61804  &  1152.69458	& 1152.84399  &  1152.84412  &  1154.77832  &  1.042-13     \\
  45  &   1s5f    &  $^1$F$_3^o$ &  	        &  1154.61816  &  1152.69702	& 1152.84656  &  1152.84656  &  1154.77844  &  1.042-13     \\
  46  &   1s5g    &  $^3$G$_3  $ &  	        &  1154.61816  &  1152.69678	& 1152.84595  &  1152.84595  &  1154.77722  &  1.747-13     \\
  47  &   1s5g    &  $^3$G$_4  $ &  	        &  1154.61816  &  1152.69495	& 1152.84412  &  1152.84412  &  1154.77722  &  1.747-13     \\
  48  &   1s5g    &  $^3$G$_5  $ &  	        &  1154.64673  &  1152.72363	& 1152.87280  &  1152.87280  &  1154.80652  &  1.753-13     \\
  49  &   1s5g    &  $^1$G$_4  $ &  	        &  1154.64673  &  1152.72510	& 1152.87427  &  1152.87427  &  1154.80652  &  1.753-13     \\
\hline 													     
\end{tabular}      
			      
%\vspace*{0.5 cm}
\begin{flushleft}
{\small
NIST: {\tt http://nist.gov/pml/data/asd.cfm} \\
GRASP1: Energies from the {\sc grasp} code with 49 level calculations {\em without} Breit and QED effects \\
GRASP2: Energies from the {\sc grasp} code with 49 level calculations {\em with} Breit and QED effects \\
FAC1: Energies from the {\sc fac} code with 49 level calculations \\
FAC2: Energies from the {\sc fac} code with 71 level calculations \\
AS: Energies from the {\sc as} code with 49 level calculations

}
\end{flushleft}
\end{table*} 

%\end{document}

\clearpage

\setcounter{table}{2} 

\begin{table*}  
\caption{a. Collision strengths for transitions in  Ga XXX. ($a{\pm}b \equiv$ $a\times$10$^{{\pm}b}$).}       
\begin{tabular}{rrllllll}                                                                                     
\hline                                                                                                        
\hline                                                                                                        
\multicolumn{2}{c}{Transition} & \multicolumn{5}{c}{Energy (Ryd)}\\                                           
\hline                                                                                                        
  $i$ & $j$ &    900 &  1200  & 1500 & 1800  &  2100  &     FAC$^a$ \\                                            
\hline   
   1 &  2 &  2.038$-$04 &  1.462$-$04 &  1.094$-$04 &  8.475$-$05 &  6.927$-$05 &  7.760$-$05 \\
  1 &  3 &  1.182$-$04 &  7.456$-$05 &  5.071$-$05 &  3.585$-$05 &  2.683$-$05 &  3.192$-$05 \\
  1 &  4 &  6.395$-$04 &  7.418$-$04 &  8.145$-$04 &  8.656$-$04 &  9.152$-$04 &  7.652$-$04 \\
  1 &  5 &  6.197$-$04 &  6.550$-$04 &  7.202$-$04 &  7.941$-$04 &  8.706$-$04 &  9.110$-$04 \\
  1 &  6 &  5.386$-$04 &  3.352$-$04 &  2.256$-$04 &  1.580$-$04 &  1.173$-$04 &  1.506$-$04 \\
  1 &  7 &  2.084$-$03 &  3.016$-$03 &  3.821$-$03 &  4.535$-$03 &  5.174$-$03 &  4.637$-$03 \\
  1 &  8 &  6.533$-$05 &  4.557$-$05 &  3.321$-$05 &  2.534$-$05 &  2.039$-$05 &  2.073$-$05 \\
  1 &  9 &  3.945$-$05 &  2.432$-$05 &  1.621$-$05 &  1.130$-$05 &  8.370$-$06 &  8.770$-$06 \\
  1 & 10 &  1.091$-$04 &  1.353$-$04 &  1.526$-$04 &  1.650$-$04 &  1.766$-$04 &  1.489$-$04 \\
  1 & 11 &  1.474$-$04 &  1.387$-$04 &  1.433$-$04 &  1.528$-$04 &  1.648$-$04 &  1.779$-$04 \\
  1 & 12 &  1.828$-$04 &  1.111$-$04 &  7.319$-$05 &  5.052$-$05 &  3.712$-$05 &  4.176$-$05 \\
  1 & 13 &  1.558$-$05 &  7.880$-$06 &  4.534$-$06 &  2.884$-$06 &  1.957$-$06 &  2.191$-$06 \\
  1 & 14 &  2.396$-$05 &  2.099$-$05 &  2.252$-$05 &  2.507$-$05 &  2.771$-$05 &  2.753$-$05 \\
  1 & 15 &  3.077$-$04 &  4.836$-$04 &  6.347$-$04 &  7.669$-$04 &  8.841$-$04 &  8.307$-$04 \\
  1 & 16 &  3.368$-$05 &  1.677$-$05 &  9.531$-$06 &  6.006$-$06 &  4.044$-$06 &  4.958$-$06 \\
  1 & 17 &  2.192$-$05 &  2.724$-$05 &  3.467$-$05 &  4.162$-$05 &  4.773$-$05 &  4.435$-$05 \\
  1 & 18 &  2.927$-$05 &  1.951$-$05 &  1.406$-$05 &  1.084$-$05 &  8.699$-$06 &  8.385$-$06 \\
  1 & 19 &  1.735$-$05 &  1.058$-$05 &  6.983$-$06 &  4.882$-$06 &  3.594$-$06 &  3.589$-$06 \\
  1 & 20 &  3.851$-$05 &  4.858$-$05 &  5.527$-$05 &  6.107$-$05 &  6.553$-$05 &  5.526$-$05 \\
  1 & 21 &  5.969$-$05 &  5.373$-$05 &  5.413$-$05 &  5.700$-$05 &  6.095$-$05 &  6.590$-$05 \\
  1 & 22 &  8.097$-$05 &  4.862$-$05 &  3.173$-$05 &  2.195$-$05 &  1.602$-$05 &  1.713$-$05 \\
  1 & 23 &  9.235$-$06 &  4.567$-$06 &  2.603$-$06 &  1.654$-$06 &  1.115$-$06 &  1.186$-$06 \\
  1 & 24 &  1.337$-$05 &  1.032$-$05 &  1.047$-$05 &  1.144$-$05 &  1.258$-$05 &  1.276$-$05 \\
  1 & 25 &  1.032$-$04 &  1.682$-$04 &  2.239$-$04 &  2.724$-$04 &  3.152$-$04 &  2.989$-$04 \\
  1 & 26 &  2.005$-$05 &  9.758$-$06 &  5.490$-$06 &  3.453$-$06 &  2.310$-$06 &  2.688$-$06 \\
  1 & 27 &  1.134$-$05 &  1.274$-$05 &  1.608$-$05 &  1.943$-$05 &  2.246$-$05 &  2.135$-$05 \\
  1 & 28 &  6.393$-$07 &  2.494$-$07 &  1.227$-$07 &  6.916$-$08 &  4.287$-$08 &  4.771$-$08 \\
  1 & 29 &  7.822$-$07 &  4.054$-$07 &  3.355$-$07 &  3.273$-$07 &  3.337$-$07 &  3.244$-$07 \\
  1 & 30 &  1.073$-$06 &  4.116$-$07 &  2.004$-$07 &  1.121$-$07 &  6.910$-$08 &  8.406$-$08 \\
  1 & 31 &  7.281$-$07 &  4.202$-$07 &  3.926$-$07 &  4.100$-$07 &  4.331$-$07 &  3.999$-$07 \\
  1 & 32 &  1.651$-$05 &  1.005$-$05 &  7.175$-$06 &  5.503$-$06 &  4.410$-$06 &  4.171$-$06 \\
  1 & 33 &  9.155$-$06 &  5.482$-$06 &  3.603$-$06 &  2.501$-$06 &  1.847$-$06 &  1.802$-$06 \\
  1 & 34 &  1.911$-$05 &  2.358$-$05 &  2.688$-$05 &  2.969$-$05 &  3.208$-$05 &  2.697$-$05 \\
  1 & 35 &  3.060$-$05 &  2.670$-$05 &  2.659$-$05 &  2.779$-$05 &  2.965$-$05 &  3.193$-$05 \\
  1 & 36 &  4.291$-$05 &  2.528$-$05 &  1.643$-$05 &  1.128$-$05 &  8.260$-$06 &  8.596$-$06 \\
  1 & 37 &  5.541$-$06 &  2.637$-$06 &  1.498$-$06 &  9.478$-$07 &  6.393$-$07 &  6.644$-$07 \\
  1 & 38 &  8.011$-$06 &  5.600$-$06 &  5.518$-$06 &  5.959$-$06 &  6.526$-$06 &  6.675$-$06 \\
  1 & 39 &  4.896$-$05 &  8.006$-$05 &  1.072$-$04 &  1.306$-$04 &  1.514$-$04 &  1.434$-$04 \\
  1 & 40 &  1.208$-$05 &  5.643$-$06 &  3.165$-$06 &  1.982$-$06 &  1.326$-$06 &  1.506$-$06 \\
  1 & 41 &  6.743$-$06 &  6.724$-$06 &  8.415$-$06 &  1.018$-$05 &  1.180$-$05 &  1.135$-$05 \\
  1 & 42 &  5.849$-$07 &  2.078$-$07 &  1.019$-$07 &  5.733$-$08 &  3.552$-$08 &  3.871$-$08 \\
  1 & 43 &  7.338$-$07 &  3.271$-$07 &  2.602$-$07 &  2.501$-$07 &  2.556$-$07 &  2.496$-$07 \\
  1 & 44 &  9.845$-$07 &  3.434$-$07 &  1.667$-$07 &  9.302$-$08 &  5.729$-$08 &  6.823$-$08 \\
  1 & 45 &  6.939$-$07 &  3.359$-$07 &  3.016$-$07 &  3.121$-$07 &  3.317$-$07 &  3.081$-$07 \\
  1 & 46 &  1.956$-$08 &  4.033$-$09 &  1.614$-$09 &  7.745$-$10 &  4.241$-$10 &  4.864$-$10 \\
  1 & 47 &  2.461$-$08 &  5.978$-$09 &  3.393$-$09 &  2.663$-$09 &  2.598$-$09 &  2.091$-$09 \\
  1 & 48 &  2.893$-$08 &  5.825$-$09 &  2.307$-$09 &  1.098$-$09 &  5.984$-$10 &  7.523$-$10 \\
  1 & 49 &  2.451$-$08 &  6.146$-$09 &  3.735$-$09 &  3.108$-$09 &  3.144$-$09 &  2.399$-$09 \\
\hline                                                                                                        
\end{tabular}    
\begin{flushleft}
$^a$ E $\sim$ 2000 Ryd
\end{flushleft}                                                                                             
\end{table*}

\clearpage  
\setcounter{table}{2} 

\begin{table*}  
\caption{b. Collision strengths for transitions in  Ge XXXI. ($a{\pm}b \equiv$ $a\times$10$^{{\pm}b}$).}      
\begin{tabular}{rrlllllll}                                                                                    
\hline                                                                                                        
\hline                                                                                                        
\multicolumn{2}{c}{Transition} & \multicolumn{7}{c}{Energy (Ryd)}\\                                           
\hline                                                                                                        
  $i$ & $j$ &    1000 &  1250  & 1500 & 1750  &  2000 &  2250  &    FAC$^a$ \\                                    
\hline   
  1 &  2 &  1.831$-$04 &  1.414$-$04 &  1.124$-$04 &  9.138$-$05 &  7.501$-$05 &  6.333$-$05 &  7.219$-$05 \\
  1 &  3 &  1.047$-$04 &  7.323$-$05 &  5.374$-$05 &  4.027$-$05 &  3.147$-$05 &  2.527$-$05 &  2.965$-$05 \\
  1 &  4 &  6.151$-$04 &  6.927$-$04 &  7.514$-$04 &  7.937$-$04 &  8.305$-$04 &  8.672$-$04 &  7.202$-$04 \\
  1 &  5 &  6.075$-$04 &  6.444$-$04 &  6.990$-$04 &  7.585$-$04 &  8.224$-$04 &  8.855$-$04 &  9.258$-$04 \\
  1 &  6 &  4.735$-$04 &  3.272$-$04 &  2.380$-$04 &  1.770$-$04 &  1.371$-$04 &  1.094$-$04 &  1.393$-$04 \\
  1 &  7 &  2.049$-$03 &  2.719$-$03 &  3.320$-$03 &  3.862$-$03 &  4.363$-$03 &  4.822$-$03 &  4.331$-$03 \\
  1 &  8 &  5.860$-$05 &  4.390$-$05 &  3.418$-$05 &  2.766$-$05 &  2.256$-$05 &  1.905$-$05 &  1.930$-$05 \\
  1 &  9 &  3.499$-$05 &  2.390$-$05 &  1.723$-$05 &  1.278$-$05 &  9.928$-$06 &  7.927$-$06 &  8.151$-$06 \\
  1 & 10 &  1.063$-$04 &  1.247$-$04 &  1.392$-$04 &  1.501$-$04 &  1.591$-$04 &  1.680$-$04 &  1.400$-$04 \\
  1 & 11 &  1.396$-$04 &  1.351$-$04 &  1.390$-$04 &  1.462$-$04 &  1.557$-$04 &  1.658$-$04 &  1.789$-$04 \\
  1 & 12 &  1.611$-$04 &  1.087$-$04 &  7.755$-$05 &  5.706$-$05 &  4.395$-$05 &  3.485$-$05 &  3.868$-$05 \\
  1 & 13 &  1.347$-$05 &  7.866$-$06 &  5.021$-$06 &  3.424$-$06 &  2.461$-$06 &  1.833$-$06 &  2.023$-$06 \\
  1 & 14 &  2.175$-$05 &  1.977$-$05 &  2.062$-$05 &  2.240$-$05 &  2.439$-$05 &  2.637$-$05 &  2.629$-$05 \\
  1 & 15 &  3.088$-$04 &  4.345$-$04 &  5.470$-$04 &  6.474$-$04 &  7.396$-$04 &  8.235$-$04 &  7.757$-$04 \\
  1 & 16 &  2.889$-$05 &  1.667$-$05 &  1.052$-$05 &  7.110$-$06 &  5.075$-$06 &  3.755$-$06 &  4.569$-$06 \\
  1 & 17 &  2.082$-$05 &  2.481$-$05 &  3.016$-$05 &  3.549$-$05 &  4.037$-$05 &  4.475$-$05 &  4.174$-$05 \\
  1 & 18 &  2.584$-$05 &  1.901$-$05 &  1.459$-$05 &  1.178$-$05 &  9.532$-$06 &  7.895$-$06 &  7.806$-$06 \\
  1 & 19 &  1.538$-$05 &  1.045$-$05 &  7.468$-$06 &  5.527$-$06 &  4.241$-$06 &  3.369$-$06 &  3.336$-$06 \\
  1 & 20 &  3.757$-$05 &  4.509$-$05 &  5.034$-$05 &  5.473$-$05 &  5.836$-$05 &  6.156$-$05 &  5.194$-$05 \\
  1 & 21 &  5.589$-$05 &  5.229$-$05 &  5.262$-$05 &  5.467$-$05 &  5.760$-$05 &  6.109$-$05 &  6.604$-$05 \\
  1 & 22 &  7.133$-$05 &  4.784$-$05 &  3.385$-$05 &  2.484$-$05 &  1.889$-$05 &  1.491$-$05 &  1.587$-$05 \\
  1 & 23 &  7.928$-$06 &  4.581$-$06 &  2.897$-$06 &  1.969$-$06 &  1.404$-$06 &  1.044$-$06 &  1.096$-$06 \\
  1 & 24 &  1.183$-$05 &  9.820$-$06 &  9.727$-$06 &  1.032$-$05 &  1.113$-$05 &  1.199$-$05 &  1.221$-$05 \\
  1 & 25 &  1.045$-$04 &  1.507$-$04 &  1.924$-$04 &  2.296$-$04 &  2.629$-$04 &  2.939$-$04 &  2.790$-$04 \\
  1 & 26 &  1.709$-$05 &  9.743$-$06 &  6.092$-$06 &  4.100$-$06 &  2.904$-$06 &  2.144$-$06 &  2.479$-$06 \\
  1 & 27 &  1.044$-$05 &  1.166$-$05 &  1.399$-$05 &  1.648$-$05 &  1.886$-$05 &  2.102$-$05 &  2.008$-$05 \\
  1 & 28 &  5.201$-$07 &  2.547$-$07 &  1.423$-$07 &  8.748$-$08 &  5.750$-$08 &  3.988$-$08 &  4.383$-$08 \\
  1 & 29 &  6.404$-$07 &  3.943$-$07 &  3.244$-$07 &  3.072$-$07 &  3.074$-$07 &  3.131$-$07 &  3.063$-$07 \\
  1 & 30 &  8.663$-$07 &  4.186$-$07 &  2.318$-$07 &  1.414$-$07 &  9.238$-$08 &  6.380$-$08 &  7.712$-$08 \\
  1 & 31 &  5.965$-$07 &  4.021$-$07 &  3.668$-$07 &  3.727$-$07 &  3.893$-$07 &  4.067$-$07 &  3.776$-$07 \\
  1 & 32 &  1.387$-$05 &  9.845$-$06 &  7.449$-$06 &  6.018$-$06 &  4.923$-$06 &  4.044$-$06 &  3.884$-$06 \\
  1 & 33 &  8.123$-$06 &  5.404$-$06 &  3.844$-$06 &  2.834$-$06 &  2.193$-$06 &  1.734$-$06 &  1.676$-$06 \\
  1 & 34 &  1.844$-$05 &  2.175$-$05 &  2.443$-$05 &  2.657$-$05 &  2.858$-$05 &  2.998$-$05 &  2.534$-$05 \\
  1 & 35 &  2.852$-$05 &  2.591$-$05 &  2.580$-$05 &  2.665$-$05 &  2.805$-$05 &  2.968$-$05 &  3.194$-$05 \\
  1 & 36 &  3.780$-$05 &  2.483$-$05 &  1.748$-$05 &  1.278$-$05 &  9.797$-$06 &  7.695$-$06 &  7.971$-$06 \\
  1 & 37 &  4.664$-$06 &  2.640$-$06 &  1.664$-$06 &  1.128$-$06 &  8.066$-$07 &  5.974$-$07 &  6.140$-$07 \\
  1 & 38 &  6.825$-$06 &  5.342$-$06 &  5.157$-$06 &  5.400$-$06 &  5.796$-$06 &  6.226$-$06 &  6.393$-$06 \\
  1 & 39 &  4.961$-$05 &  7.157$-$05 &  9.171$-$05 &  1.098$-$04 &  1.260$-$04 &  1.412$-$04 &  1.339$-$04 \\
  1 & 40 &  1.007$-$05 &  5.625$-$06 &  3.504$-$06 &  2.354$-$06 &  1.670$-$06 &  1.229$-$06 &  1.389$-$06 \\
  1 & 41 &  5.856$-$06 &  6.161$-$06 &  7.328$-$06 &  8.630$-$06 &  9.893$-$06 &  1.105$-$05 &  1.066$-$05 \\
  1 & 42 &  4.407$-$07 &  2.119$-$07 &  1.182$-$07 &  7.255$-$08 &  4.768$-$08 &  3.301$-$08 &  3.558$-$08 \\
  1 & 43 &  5.537$-$07 &  3.193$-$07 &  2.537$-$07 &  2.358$-$07 &  2.352$-$07 &  2.400$-$07 &  2.356$-$07 \\
  1 & 44 &  7.352$-$07 &  3.489$-$07 &  1.927$-$07 &  1.174$-$07 &  7.666$-$08 &  5.285$-$08 &  6.261$-$08 \\
  1 & 45 &  5.198$-$07 &  3.228$-$07 &  2.837$-$07 &  2.842$-$07 &  2.970$-$07 &  3.117$-$07 &  2.909$-$07 \\
  1 & 46 &  1.154$-$08 &  4.197$-$09 &  1.985$-$09 &  1.060$-$09 &  6.229$-$10 &  3.943$-$10 &  4.442$-$10 \\
  1 & 47 &  1.515$-$08 &  6.024$-$09 &  3.629$-$09 &  2.724$-$09 &  2.414$-$09 &  2.385$-$09 &  1.970$-$09 \\
  1 & 48 &  1.683$-$08 &  6.037$-$09 &  2.828$-$09 &  1.499$-$09 &  8.759$-$10 &  5.524$-$10 &  6.862$-$10 \\
  1 & 49 &  1.512$-$08 &  6.141$-$09 &  3.900$-$09 &  3.086$-$09 &  2.847$-$09 &  2.889$-$09 &  2.263$-$09 \\
  2 &  3 &  8.169$-$02 &  7.548$-$02 &  7.665$-$02 &  7.494$-$02 &  7.413$-$02 &  7.150$-$02 &  1.268$-$01 \\
\hline                                                                                                        
\end{tabular}    
\begin{flushleft}
$^a$ E $\sim$ 2100 Ryd
\end{flushleft}                                                                                              
\end{table*}                                                                       
                                                       
\clearpage  
\setcounter{table}{2} 

\begin{table*}  
\caption{c. Collision strengths for transitions in  As XXXII. ($a{\pm}b \equiv$ $a\times$10$^{{\pm}b}$).}     
\begin{tabular}{rrlllllll}                                                                                    
\hline                                                                                                        
\hline                                                                                                        
\multicolumn{2}{c}{Transition} & \multicolumn{7}{c}{Energy (Ryd)}\\                                           
\hline                                                                                                        
  $i$ & $j$ &    1100 &  1400  & 1700 & 2000  &  2300 &  2600  &    FAC$^a$ \\                                    
\hline 
  1 &  2 &  1.681$-$04 &  1.254$-$04 &  9.695$-$05 &  7.778$-$05 &  6.367$-$05 &  5.333$-$05 &  6.734$-$05 \\
  1 &  3 &  9.376$-$05 &  6.353$-$05 &  4.526$-$05 &  3.359$-$05 &  2.589$-$05 &  2.048$-$05 &  2.763$-$05 \\
  1 &  4 &  5.950$-$04 &  6.689$-$04 &  7.265$-$04 &  7.704$-$04 &  8.122$-$04 &  8.460$-$04 &  6.795$-$04 \\
  1 &  5 &  5.975$-$04 &  6.516$-$04 &  7.211$-$04 &  7.947$-$04 &  8.697$-$04 &  9.419$-$04 &  9.364$-$04 \\
  1 &  6 &  4.205$-$04 &  2.816$-$04 &  1.981$-$04 &  1.458$-$04 &  1.113$-$04 &  8.736$-$05 &  1.293$-$04 \\
  1 &  7 &  1.992$-$03 &  2.675$-$03 &  3.286$-$03 &  3.833$-$03 &  4.335$-$03 &  4.794$-$03 &  4.052$-$03 \\
  1 &  8 &  5.347$-$05 &  3.897$-$05 &  2.955$-$05 &  2.325$-$05 &  1.921$-$05 &  1.582$-$05 &  1.801$-$05 \\
  1 &  9 &  3.115$-$05 &  2.072$-$05 &  1.448$-$05 &  1.060$-$05 &  8.135$-$06 &  6.376$-$06 &  7.599$-$06 \\
  1 & 10 &  1.035$-$04 &  1.222$-$04 &  1.356$-$04 &  1.459$-$04 &  1.566$-$04 &  1.647$-$04 &  1.320$-$04 \\
  1 & 11 &  1.325$-$04 &  1.325$-$04 &  1.396$-$04 &  1.499$-$04 &  1.618$-$04 &  1.735$-$04 &  1.792$-$04 \\
  1 & 12 &  1.423$-$04 &  9.339$-$05 &  6.449$-$05 &  4.675$-$05 &  3.557$-$05 &  2.764$-$05 &  3.594$-$05 \\
  1 & 13 &  1.182$-$05 &  6.580$-$06 &  4.057$-$06 &  2.710$-$06 &  1.921$-$06 &  1.408$-$06 &  1.874$-$06 \\
  1 & 14 &  1.995$-$05 &  1.874$-$05 &  2.009$-$05 &  2.218$-$05 &  2.435$-$05 &  2.642$-$05 &  2.512$-$05 \\
  1 & 15 &  3.034$-$04 &  4.325$-$04 &  5.459$-$04 &  6.471$-$04 &  7.389$-$04 &  8.214$-$04 &  7.259$-$04 \\
  1 & 16 &  2.520$-$05 &  1.381$-$05 &  8.411$-$06 &  5.565$-$06 &  3.909$-$06 &  2.848$-$06 &  4.224$-$06 \\
  1 & 17 &  1.979$-$05 &  2.447$-$05 &  3.015$-$05 &  3.561$-$05 &  4.046$-$05 &  4.480$-$05 &  3.937$-$05 \\
  1 & 18 &  2.360$-$05 &  1.668$-$05 &  1.268$-$05 &  9.956$-$06 &  7.970$-$06 &  6.608$-$06 &  7.287$-$06 \\
  1 & 19 &  1.378$-$05 &  9.002$-$06 &  6.294$-$06 &  4.591$-$06 &  3.461$-$06 &  2.722$-$06 &  3.111$-$06 \\
  1 & 20 &  3.696$-$05 &  4.382$-$05 &  4.974$-$05 &  5.388$-$05 &  5.730$-$05 &  6.055$-$05 &  4.893$-$05 \\
  1 & 21 &  5.272$-$05 &  5.058$-$05 &  5.231$-$05 &  5.557$-$05 &  5.949$-$05 &  6.373$-$05 &  6.594$-$05 \\
  1 & 22 &  6.341$-$05 &  4.086$-$05 &  2.822$-$05 &  2.038$-$05 &  1.524$-$05 &  1.188$-$05 &  1.476$-$05 \\
  1 & 23 &  6.987$-$06 &  3.809$-$06 &  2.344$-$06 &  1.557$-$06 &  1.095$-$06 &  8.022$-$07 &  1.016$-$06 \\
  1 & 24 &  1.074$-$05 &  9.140$-$06 &  9.388$-$06 &  1.018$-$05 &  1.110$-$05 &  1.204$-$05 &  1.169$-$05 \\
  1 & 25 &  1.035$-$04 &  1.507$-$04 &  1.923$-$04 &  2.296$-$04 &  2.634$-$04 &  2.940$-$04 &  2.611$-$04 \\
  1 & 26 &  1.495$-$05 &  8.026$-$06 &  4.879$-$06 &  3.208$-$06 &  2.237$-$06 &  1.628$-$06 &  2.292$-$06 \\
  1 & 27 &  9.809$-$06 &  1.142$-$05 &  1.398$-$05 &  1.658$-$05 &  1.895$-$05 &  2.111$-$05 &  1.892$-$05 \\
  1 & 28 &  4.418$-$07 &  2.046$-$07 &  1.105$-$07 &  6.629$-$08 &  4.277$-$08 &  2.937$-$08 &  4.040$-$08 \\
  1 & 29 &  5.510$-$07 &  3.440$-$07 &  2.951$-$07 &  2.876$-$07 &  2.917$-$07 &  3.004$-$07 &  2.897$-$07 \\
  1 & 30 &  7.303$-$07 &  3.334$-$07 &  1.781$-$07 &  1.060$-$07 &  6.802$-$08 &  4.649$-$08 &  7.097$-$08 \\
  1 & 31 &  5.148$-$07 &  3.601$-$07 &  3.438$-$07 &  3.575$-$07 &  3.761$-$07 &  3.953$-$07 &  3.571$-$07 \\
  1 & 32 &  1.241$-$05 &  8.669$-$06 &  6.412$-$06 &  5.071$-$06 &  4.084$-$06 &  3.346$-$06 &  3.627$-$06 \\
  1 & 33 &  7.165$-$06 &  4.656$-$06 &  3.229$-$06 &  2.350$-$06 &  1.782$-$06 &  1.392$-$06 &  1.563$-$06 \\
  1 & 34 &  1.775$-$05 &  2.118$-$05 &  2.406$-$05 &  2.610$-$05 &  2.788$-$05 &  2.955$-$05 &  2.386$-$05 \\
  1 & 35 &  2.643$-$05 &  2.493$-$05 &  2.552$-$05 &  2.699$-$05 &  2.887$-$05 &  3.083$-$05 &  3.185$-$05 \\
  1 & 36 &  3.312$-$05 &  2.121$-$05 &  1.453$-$05 &  1.047$-$05 &  7.868$-$06 &  6.091$-$06 &  7.412$-$06 \\
  1 & 37 &  4.045$-$06 &  2.196$-$06 &  1.343$-$06 &  8.921$-$07 &  6.269$-$07 &  4.580$-$07 &  5.690$-$07 \\
  1 & 38 &  6.060$-$06 &  4.933$-$06 &  4.947$-$06 &  5.315$-$06 &  5.775$-$06 &  6.252$-$06 &  6.125$-$06 \\
  1 & 39 &  4.879$-$05 &  7.153$-$05 &  9.179$-$05 &  1.099$-$04 &  1.264$-$04 &  1.411$-$04 &  1.252$-$04 \\
  1 & 40 &  8.676$-$06 &  4.634$-$06 &  2.802$-$06 &  1.841$-$06 &  1.282$-$06 &  9.309$-$07 &  1.285$-$06 \\
  1 & 41 &  5.366$-$06 &  6.015$-$06 &  7.316$-$06 &  8.686$-$06 &  9.950$-$06 &  1.111$-$05 &  1.005$-$05 \\
  1 & 42 &  3.673$-$07 &  1.702$-$07 &  9.168$-$08 &  5.498$-$08 &  3.543$-$08 &  2.432$-$08 &  3.280$-$08 \\
  1 & 43 &  4.638$-$07 &  2.758$-$07 &  2.287$-$07 &  2.199$-$07 &  2.234$-$07 &  2.309$-$07 &  2.229$-$07 \\
  1 & 44 &  6.081$-$07 &  2.778$-$07 &  1.480$-$07 &  8.803$-$08 &  5.639$-$08 &  3.854$-$08 &  5.764$-$08 \\
  1 & 45 &  4.351$-$07 &  2.856$-$07 &  2.639$-$07 &  2.720$-$07 &  2.876$-$07 &  3.041$-$07 &  2.752$-$07 \\
  1 & 46 &  8.920$-$09 &  3.219$-$09 &  1.456$-$09 &  7.558$-$10 &  4.362$-$10 &  2.742$-$10 &  4.070$-$10 \\
  1 & 47 &  1.173$-$08 &  4.819$-$09 &  2.966$-$09 &  2.342$-$09 &  2.195$-$09 &  2.240$-$09 &  1.853$-$09 \\
  1 & 48 &  1.291$-$08 &  4.591$-$09 &  2.054$-$09 &  1.058$-$09 &  6.073$-$10 &  3.803$-$10 &  6.281$-$10 \\
  1 & 49 &  1.170$-$08 &  4.963$-$09 &  3.250$-$09 &  2.711$-$09 &  2.637$-$09 &  2.748$-$09 &  2.131$-$09 \\
\hline                                                                                                        
\end{tabular}     
\begin{flushleft}
$^a$ E $\sim$ 2300 Ryd
\end{flushleft}                                                                                             
\end{table*}                                                                                

\clearpage  
\setcounter{table}{2} 

\begin{table*}  
\caption{d. Collision strengths for transitions in  Se XXXIII. ($a{\pm}b \equiv$ $a\times$10$^{{\pm}b}$).}    
\begin{tabular}{rrllllll}                                                                                     
\hline                                                                                                        
\hline                                                                                                        
\multicolumn{2}{c}{Transition} & \multicolumn{6}{c}{Energy (Ryd)}\\                                           
\hline                                                                                                        
  $i$ & $j$ &    1100 &  1500  & 2000 & 2500  &  3000   &    FAC$^a$ \\                                           
\hline   
  1 &  2 &  1.688$-$04 &  1.178$-$04 &  8.198$-$05 &  5.882$-$05 &  4.510$-$05 &  6.298$-$05 \\
  1 &  3 &  9.769$-$05 &  5.972$-$05 &  3.578$-$05 &  2.360$-$05 &  1.654$-$05 &  2.582$-$05 \\
  1 &  4 &  5.430$-$04 &  6.415$-$04 &  7.215$-$04 &  7.767$-$04 &  8.292$-$04 &  6.421$-$04 \\
  1 &  5 &  5.752$-$04 &  6.483$-$04 &  7.653$-$04 &  8.856$-$04 &  1.001$-$03 &  9.430$-$04 \\
  1 &  6 &  4.367$-$04 &  2.620$-$04 &  1.545$-$04 &  1.005$-$04 &  6.952$-$05 &  1.203$-$04 \\
  1 &  7 &  1.705$-$03 &  2.516$-$03 &  3.378$-$03 &  4.119$-$03 &  4.775$-$03 &  3.799$-$03 \\
  1 &  8 &  5.414$-$05 &  3.627$-$05 &  2.472$-$05 &  1.736$-$05 &  1.347$-$05 &  1.685$-$05 \\
  1 &  9 &  3.260$-$05 &  1.932$-$05 &  1.131$-$05 &  7.350$-$06 &  5.127$-$06 &  7.104$-$06 \\
  1 & 10 &  9.269$-$05 &  1.162$-$04 &  1.357$-$04 &  1.496$-$04 &  1.619$-$04 &  1.246$-$04 \\
  1 & 11 &  1.287$-$04 &  1.292$-$04 &  1.439$-$04 &  1.628$-$04 &  1.824$-$04 &  1.789$-$04 \\
  1 & 12 &  1.487$-$04 &  8.641$-$05 &  4.973$-$05 &  3.183$-$05 &  2.191$-$05 &  3.348$-$05 \\
  1 & 13 &  1.276$-$05 &  6.077$-$06 &  2.994$-$06 &  1.718$-$06 &  1.094$-$06 &  1.741$-$06 \\
  1 & 14 &  1.964$-$05 &  1.769$-$05 &  2.023$-$05 &  2.348$-$05 &  2.655$-$05 &  2.403$-$05 \\
  1 & 15 &  2.544$-$04 &  4.068$-$04 &  5.670$-$04 &  7.028$-$04 &  8.221$-$04 &  6.807$-$04 \\
  1 & 16 &  2.713$-$05 &  1.266$-$05 &  6.120$-$06 &  3.465$-$06 &  2.183$-$06 &  3.915$-$06 \\
  1 & 17 &  1.806$-$05 &  2.311$-$05 &  3.142$-$05 &  3.874$-$05 &  4.490$-$05 &  3.722$-$05 \\
  1 & 18 &  2.407$-$05 &  1.561$-$05 &  1.039$-$05 &  7.375$-$06 &  5.598$-$06 &  6.820$-$06 \\
  1 & 19 &  1.434$-$05 &  8.413$-$06 &  4.860$-$06 &  3.161$-$06 &  2.189$-$06 &  2.909$-$06 \\
  1 & 20 &  3.266$-$05 &  4.186$-$05 &  4.954$-$05 &  5.489$-$05 &  5.991$-$05 &  4.616$-$05 \\
  1 & 21 &  5.122$-$05 &  4.912$-$05 &  5.323$-$05 &  5.976$-$05 &  6.661$-$05 &  6.563$-$05 \\
  1 & 22 &  6.594$-$05 &  3.792$-$05 &  2.152$-$05 &  1.378$-$05 &  9.420$-$06 &  1.375$-$05 \\
  1 & 23 &  7.552$-$06 &  3.529$-$06 &  1.721$-$06 &  9.803$-$07 &  6.227$-$07 &  9.439$-$07 \\
  1 & 24 &  1.084$-$05 &  8.624$-$06 &  9.335$-$06 &  1.073$-$05 &  1.214$-$05 &  1.120$-$05 \\
  1 & 25 &  8.576$-$05 &  1.418$-$04 &  2.008$-$04 &  2.508$-$04 &  2.943$-$04 &  2.449$-$04 \\
  1 & 26 &  1.615$-$05 &  7.384$-$06 &  3.531$-$06 &  1.983$-$06 &  1.246$-$06 &  2.126$-$06 \\
  1 & 27 &  9.241$-$06 &  1.078$-$05 &  1.459$-$05 &  1.816$-$05 &  2.123$-$05 &  1.787$-$05 \\
  1 & 28 &  5.086$-$07 &  1.882$-$07 &  7.589$-$08 &  3.788$-$08 &  2.180$-$08 &  3.734$-$08 \\
  1 & 29 &  6.117$-$07 &  3.194$-$07 &  2.700$-$07 &  2.762$-$07 &  2.896$-$07 &  2.744$-$07 \\
  1 & 30 &  8.312$-$07 &  3.047$-$07 &  1.209$-$07 &  5.971$-$08 &  3.415$-$08 &  6.550$-$08 \\
  1 & 31 &  5.676$-$07 &  3.352$-$07 &  3.284$-$07 &  3.577$-$07 &  3.851$-$07 &  3.383$-$07 \\
  1 & 32 &  1.323$-$05 &  8.043$-$06 &  5.343$-$06 &  3.761$-$06 &  2.820$-$06 &  3.395$-$06 \\
  1 & 33 &  7.637$-$06 &  4.383$-$06 &  2.504$-$06 &  1.622$-$06 &  1.116$-$06 &  1.462$-$06 \\
  1 & 34 &  1.613$-$05 &  2.040$-$05 &  2.400$-$05 &  2.667$-$05 &  2.912$-$05 &  2.249$-$05 \\
  1 & 35 &  2.628$-$05 &  2.423$-$05 &  2.587$-$05 &  2.895$-$05 &  3.216$-$05 &  3.166$-$05 \\
  1 & 36 &  3.528$-$05 &  1.983$-$05 &  1.113$-$05 &  7.098$-$06 &  4.822$-$06 &  6.911$-$06 \\
  1 & 37 &  4.516$-$06 &  2.043$-$06 &  9.869$-$07 &  5.620$-$07 &  3.553$-$07 &  5.289$-$07 \\
  1 & 38 &  6.378$-$06 &  4.659$-$06 &  4.893$-$06 &  5.582$-$06 &  6.310$-$06 &  5.871$-$06 \\
  1 & 39 &  4.088$-$05 &  6.737$-$05 &  9.596$-$05 &  1.204$-$04 &  1.413$-$04 &  1.174$-$04 \\
  1 & 40 &  9.691$-$06 &  4.283$-$06 &  2.028$-$06 &  1.139$-$06 &  7.126$-$07 &  1.192$-$06 \\
  1 & 41 &  5.344$-$06 &  5.680$-$06 &  7.638$-$06 &  9.531$-$06 &  1.118$-$05 &  9.488$-$06 \\
  1 & 42 &  4.472$-$07 &  1.572$-$07 &  6.291$-$08 &  3.141$-$08 &  1.806$-$08 &  3.033$-$08 \\
  1 & 43 &  5.451$-$07 &  2.570$-$07 &  2.075$-$07 &  2.114$-$07 &  2.235$-$07 &  2.112$-$07 \\
  1 & 44 &  7.171$-$07 &  2.549$-$07 &  1.003$-$07 &  4.956$-$08 &  2.830$-$08 &  5.321$-$08 \\
  1 & 45 &  5.211$-$07 &  2.667$-$07 &  2.506$-$07 &  2.736$-$07 &  2.977$-$07 &  2.607$-$07 \\
  1 & 46 &  1.292$-$08 &  2.950$-$09 &  9.123$-$10 &  3.792$-$10 &  1.922$-$10 &  3.741$-$10 \\
  1 & 47 &  1.662$-$08 &  4.415$-$09 &  2.344$-$09 &  2.031$-$09 &  2.138$-$09 &  1.746$-$09 \\
  1 & 48 &  1.848$-$08 &  4.179$-$09 &  1.272$-$09 &  5.236$-$10 &  2.640$-$10 &  5.766$-$10 \\
  1 & 49 &  1.612$-$08 &  4.545$-$09 &  2.659$-$09 &  2.451$-$09 &  2.649$-$09 &  2.010$-$09 \\
\hline                                                                                                        
\end{tabular}     
\begin{flushleft}
$^a$ E $\sim$ 2400 Ryd
\end{flushleft}                                                                                             
\end{table*}                                                                                              

\clearpage  
\setcounter{table}{2} 

\begin{table*}  
\caption{e. Collision strengths for transitions in  Br XXXIV. ($a{\pm}b \equiv$ $a\times$10$^{{\pm}b}$).}     
\begin{tabular}{rrlllllll}                                                                                    
\hline                                                                                                        
\hline                                                                                                        
\multicolumn{2}{c}{Transition} & \multicolumn{7}{c}{Energy (Ryd)}\\                                           
\hline                                                                                                        
  $i$ & $j$ &    1200 &  1600  & 2000 & 2400  &  2800   &   3200 &   FAC$^a$ \\                                   
\hline    
  1 &  2 &  1.557$-$04 &  1.105$-$04 &  8.404$-$05 &  6.450$-$05 &  5.230$-$05 &  4.178$-$05 &  5.905$-$05 \\
  1 &  3 &  8.866$-$05 &  5.595$-$05 &  3.776$-$05 &  2.697$-$05 &  2.022$-$05 &  1.563$-$05 &  2.419$-$05 \\
  1 &  4 &  5.262$-$04 &  6.075$-$04 &  6.740$-$04 &  7.179$-$04 &  7.597$-$04 &  7.883$-$04 &  6.077$-$04 \\
  1 &  5 &  5.638$-$04 &  6.411$-$04 &  7.325$-$04 &  8.267$-$04 &  9.178$-$04 &  1.007$-$03 &  9.459$-$04 \\
  1 &  6 &  3.930$-$04 &  2.438$-$04 &  1.620$-$04 &  1.144$-$04 &  8.491$-$05 &  6.501$-$05 &  1.122$-$04 \\
  1 &  7 &  1.651$-$03 &  2.360$-$03 &  2.978$-$03 &  3.529$-$03 &  4.024$-$03 &  4.487$-$03 &  3.569$-$03 \\
  1 &  8 &  5.012$-$05 &  3.424$-$05 &  2.571$-$05 &  1.959$-$05 &  1.547$-$05 &  1.228$-$05 &  1.580$-$05 \\
  1 &  9 &  2.977$-$05 &  1.820$-$05 &  1.208$-$05 &  8.517$-$06 &  6.305$-$06 &  4.861$-$06 &  6.659$-$06 \\
  1 & 10 &  9.115$-$05 &  1.105$-$04 &  1.266$-$04 &  1.371$-$04 &  1.472$-$04 &  1.536$-$04 &  1.178$-$04 \\
  1 & 11 &  1.235$-$04 &  1.263$-$04 &  1.377$-$04 &  1.520$-$04 &  1.670$-$04 &  1.823$-$04 &  1.780$-$04 \\
  1 & 12 &  1.346$-$04 &  8.084$-$05 &  5.287$-$05 &  3.681$-$05 &  2.694$-$05 &  2.057$-$05 &  3.127$-$05 \\
  1 & 13 &  1.138$-$05 &  5.695$-$06 &  3.304$-$06 &  2.102$-$06 &  1.428$-$06 &  1.028$-$06 &  1.622$-$06 \\
  1 & 14 &  1.819$-$05 &  1.677$-$05 &  1.849$-$05 &  2.083$-$05 &  2.315$-$05 &  2.538$-$05 &  2.300$-$05 \\
  1 & 15 &  2.505$-$04 &  3.830$-$04 &  4.977$-$04 &  5.990$-$04 &  6.894$-$04 &  7.736$-$04 &  6.398$-$04 \\
  1 & 16 &  2.403$-$05 &  1.178$-$05 &  6.729$-$06 &  4.226$-$06 &  2.846$-$06 &  2.032$-$06 &  3.638$-$06 \\
  1 & 17 &  1.722$-$05 &  2.182$-$05 &  2.774$-$05 &  3.327$-$05 &  3.812$-$05 &  4.247$-$05 &  3.525$-$05 \\
  1 & 18 &  2.226$-$05 &  1.483$-$05 &  1.082$-$05 &  8.152$-$06 &  6.502$-$06 &  5.091$-$06 &  6.398$-$06 \\
  1 & 19 &  1.322$-$05 &  7.955$-$06 &  5.176$-$06 &  3.631$-$06 &  2.688$-$06 &  2.058$-$06 &  2.727$-$06 \\
  1 & 20 &  3.237$-$05 &  4.003$-$05 &  4.581$-$05 &  5.006$-$05 &  5.393$-$05 &  5.643$-$05 &  4.363$-$05 \\
  1 & 21 &  4.906$-$05 &  4.780$-$05 &  5.084$-$05 &  5.571$-$05 &  6.090$-$05 &  6.635$-$05 &  6.513$-$05 \\
  1 & 22 &  6.031$-$05 &  3.562$-$05 &  2.283$-$05 &  1.582$-$05 &  1.157$-$05 &  8.773$-$06 &  1.285$-$05 \\
  1 & 23 &  6.766$-$06 &  3.317$-$06 &  1.893$-$06 &  1.202$-$06 &  8.122$-$07 &  5.840$-$07 &  8.796$-$07 \\
  1 & 24 &  9.954$-$06 &  8.182$-$06 &  8.598$-$06 &  9.551$-$06 &  1.060$-$05 &  1.163$-$05 &  1.073$-$05 \\
  1 & 25 &  8.523$-$05 &  1.335$-$04 &  1.755$-$04 &  2.130$-$04 &  2.460$-$04 &  2.768$-$04 &  2.301$-$04 \\
  1 & 26 &  1.436$-$05 &  6.893$-$06 &  3.870$-$06 &  2.426$-$06 &  1.623$-$06 &  1.159$-$06 &  1.977$-$06 \\
  1 & 27 &  8.711$-$06 &  1.019$-$05 &  1.286$-$05 &  1.551$-$05 &  1.791$-$05 &  2.008$-$05 &  1.692$-$05 \\
  1 & 28 &  4.461$-$07 &  1.763$-$07 &  8.636$-$08 &  4.892$-$08 &  3.038$-$08 &  2.035$-$08 &  3.462$-$08 \\
  1 & 29 &  5.424$-$07 &  3.001$-$07 &  2.565$-$07 &  2.553$-$07 &  2.637$-$07 &  2.741$-$07 &  2.604$-$07 \\
  1 & 30 &  7.311$-$07 &  2.835$-$07 &  1.370$-$07 &  7.686$-$08 &  4.742$-$08 &  3.159$-$08 &  6.062$-$08 \\
  1 & 31 &  4.996$-$07 &  3.150$-$07 &  3.042$-$07 &  3.231$-$07 &  3.450$-$07 &  3.650$-$07 &  3.210$-$07 \\
  1 & 32 &  1.181$-$05 &  7.695$-$06 &  5.519$-$06 &  4.201$-$06 &  3.295$-$06 &  2.644$-$06 &  3.185$-$06 \\
  1 & 33 &  6.818$-$06 &  4.103$-$06 &  2.680$-$06 &  1.868$-$06 &  1.385$-$06 &  1.056$-$06 &  1.371$-$06 \\
  1 & 34 &  1.538$-$05 &  1.924$-$05 &  2.230$-$05 &  2.429$-$05 &  2.633$-$05 &  2.750$-$05 &  2.125$-$05 \\
  1 & 35 &  2.438$-$05 &  2.344$-$05 &  2.477$-$05 &  2.696$-$05 &  2.947$-$05 &  3.200$-$05 &  3.137$-$05 \\
  1 & 36 &  3.125$-$05 &  1.845$-$05 &  1.187$-$05 &  8.168$-$06 &  5.986$-$06 &  4.513$-$06 &  6.460$-$06 \\
  1 & 37 &  3.891$-$06 &  1.911$-$06 &  1.091$-$06 &  6.881$-$07 &  4.658$-$07 &  3.333$-$07 &  4.929$-$07 \\
  1 & 38 &  5.620$-$06 &  4.406$-$06 &  4.526$-$06 &  4.986$-$06 &  5.515$-$06 &  6.052$-$06 &  5.631$-$06 \\
  1 & 39 &  3.987$-$05 &  6.344$-$05 &  8.390$-$05 &  1.020$-$04 &  1.181$-$04 &  1.329$-$04 &  1.103$-$04 \\
  1 & 40 &  8.284$-$06 &  3.979$-$06 &  2.236$-$06 &  1.391$-$06 &  9.325$-$07 &  6.624$-$07 &  1.108$-$06 \\
  1 & 41 &  4.800$-$06 &  5.361$-$06 &  6.727$-$06 &  8.131$-$06 &  9.410$-$06 &  1.058$-$05 &  8.976$-$06 \\
  1 & 42 &  3.697$-$07 &  1.467$-$07 &  7.187$-$08 &  4.053$-$08 &  2.519$-$08 &  1.686$-$08 &  2.812$-$08 \\
  1 & 43 &  4.582$-$07 &  2.399$-$07 &  1.986$-$07 &  1.953$-$07 &  2.023$-$07 &  2.119$-$07 &  2.004$-$07 \\
  1 & 44 &  6.073$-$07 &  2.364$-$07 &  1.142$-$07 &  6.377$-$08 &  3.934$-$08 &  2.620$-$08 &  4.926$-$08 \\
  1 & 45 &  4.249$-$07 &  2.488$-$07 &  2.333$-$07 &  2.464$-$07 &  2.647$-$07 &  2.825$-$07 &  2.474$-$07 \\
  1 & 46 &  9.556$-$09 &  2.739$-$09 &  1.099$-$09 &  5.303$-$10 &  2.930$-$10 &  1.792$-$10 &  3.448$-$10 \\
  1 & 47 &  1.232$-$08 &  4.089$-$09 &  2.408$-$09 &  1.943$-$09 &  1.898$-$09 &  1.981$-$09 &  1.642$-$09 \\
  1 & 48 &  1.374$-$08 &  3.855$-$09 &  1.526$-$09 &  7.293$-$10 &  4.005$-$10 &  2.440$-$10 &  5.309$-$10 \\
  1 & 49 &  1.222$-$08 &  4.207$-$09 &  2.673$-$09 &  2.289$-$09 &  2.315$-$09 &  2.456$-$09 &  1.891$-$09 \\
\hline                                                                                                        
\end{tabular}   
\begin{flushleft}
$^a$ E $\sim$ 2600 Ryd
\end{flushleft}                                                                                               
\end{table*}                                                                            
                                                                
\end{document}